\documentclass[aps,twocolumn,superscriptaddress,notitlepage]{revtex4-2}
\usepackage{graphicx}
\usepackage[T1]{fontenc}
\usepackage{xfrac}
\usepackage{upgreek}
\usepackage[usenames,dvipsnames]{xcolor}
\usepackage[linktoc=page,colorlinks,urlcolor=OliveGreen,citecolor=OliveGreen,linkcolor=OliveGreen]{hyperref}
\usepackage{amssymb,amsmath}
\usepackage{graphicx}
\usepackage{float}
\usepackage{natbib}
\usepackage{mathrsfs}
\usepackage{overpic}
\usepackage{wrapfig}
\usepackage{braket}
\usepackage{color}
\usepackage{verbatim}
\usepackage{mathtools}
\usepackage{soul,xcolor}
\usepackage[letterpaper,textwidth=7in,top=.75in,bottom=.75in]{geometry}
\definecolor{darkblue}{rgb}{0.0 0.0 0.78}
\definecolor{darkred}{rgb}{0.5 0.0 0.0}

\newcommand{\UMDphy}{Department of Physics, University of Maryland, College Park, MD 20742, USA}
\newcommand{\QTC}{Quantum Technology Center, University of Maryland, College Park, MD 20742, USA}
\newcommand{\UMDEECS}{Department of Electrical and Computer Engineering,
University of Maryland, College Park, MD 20742, USA}

\newcommand{\YSUphy}{Department of Physics, Youngstown State University, Youngstown, Ohio}
\newcommand{\Columbia}{Departamento de Física, Universidad Nacional de Colombia, sede Bogotá, Carrera 45, Colombia}

\newcommand{\LPS}{Laboratory for Physical Sciences, 8050 Greenmead Drive, College Park, Maryland 20740, USA}

\begin{document}

\setstcolor{red}
\preprint{APS/123-QED}

\title{All-Optical Photoluminescence Response of Nitrogen-Vacancy Ensembles in Diamond at Low Magnetic Fields}

\author{Xiechen Zheng}
\thanks{These authors contributed equally to this work.}
\affiliation{\UMDEECS}
\affiliation{\QTC}
\author{Jeyson Támara-Isaza}
\thanks{These authors contributed equally to this work.}
\affiliation{\QTC}
\affiliation{\Columbia}
%\thanks{Present Address: \SQC}
\author{Zechuan Yin}
\affiliation{\UMDEECS}
\affiliation{\QTC}
\author{Johannes Cremer}
% \affiliation{\UMDEECS}
\affiliation{\QTC}
\author{John W. Blanchard}
\affiliation{\QTC}
\author{\\Connor A. Hart}
% \affiliation{\UMDEECS}
\affiliation{\QTC}
\author{Michael Crescimanno}
\affiliation{\QTC}
\affiliation{\YSUphy}
\author{Paul V. Petruzzi}
\affiliation{\LPS}
\author{Matthew J. Turner}
% \affiliation{\UMDEECS}
\affiliation{\QTC}
\author{Ronald L. Walsworth}
\affiliation{\UMDEECS}
\affiliation{\QTC}
\affiliation{\UMDphy}

\date{\today}% It is always \today, today,

\begin{abstract}

All-optical (AO), microwave-free magnetometry using nitrogen-vacancy (NV) centers in diamond is attractive due to its broad applicability and reduced experimental complexity. 
In this work, we investigate room-temperature AO photoluminescence (PL) at low magnetic fields (<2\,mT) using diamonds with NV ensembles at ppm concentrations. 
Measured AO-PL contrast features as a function of applied magnetic field magnitude and direction are correlated with near-degenerate NV electronic spin and hyperfine transitions from different NV orientations within the diamond host.
Reasonable agreement is found between low-field AO-PL measurements and model-based simulations of the effects of resonant dipolar interactions between NV centers.
Maximum observed AO-PL contrast depends on both NV concentration and laser illumination intensity at 532\,nm. 
These results imply different optimal conditions for low-field AO NV sensing compared to conventional optically detected magnetic resonance (ODMR) techniques, suggesting new research and application opportunities using AO measurements with lower system complexity, size, weight, and power.

\end{abstract}

\maketitle
\section{Introduction}
Nitrogen-vacancy (NV) centers in diamonds are a leading modality for magnetometry under wide-ranging conditions, enabling diverse applications across the physical and life sciences \cite{taylor_high-sensitivity_2008, barry_optical_2016, glenn_high-resolution_2018, ku_imaging_2020, turner_magnetic_2020, tang_quantum_2023, yin_quantum_2024}. 
Predominant NV sensing protocols, such as optically detected magnetic resonance (ODMR), require microwave fields to coherently manipulate NV electronic spin states, with magnetometry information read out via spin-state-dependent NV photoluminescence (PL). 
However, the use of external microwave fields adds design complexity and can be incompatible with the system or sample under study \cite{vorst_rf_2006, schlussel_wide-field_2018}. 

Alternatively, all-optical (AO), microwave-free NV magnetometry protocols have been demonstrated. 
AO magnetometry exploits effects that reduce PL spin-state contrast, e.g., NV spin-state mixing for off-axis magnetic fields $\sim10$\,mT \cite{lai_influence_2009, rondin_nanoscale_2012, tetienne_magnetic-field-dependent_2012}, cross-relaxation between NV electronic spins and substitutional nitrogen (P1) centers at an applied bias magnetic field near 50\,mT \cite{zhang_battery_2021}, cross-relaxation with NV centers not aligned with the bias field near 60\,mT \cite{jarmola_temperature-_2012}, and NV electronic spin-state mixing near the ground-state level anti-crossing (GSLAC) at a bias field $\approx 100$\,mT \cite{wickenbrock_microwave-free_2016}. 
These approaches to AO magnetometry employ substantial applied fields that can increase system complexity and SWaP (size, weight, and power); and also induce undesired effects, such as perturbing magnetic materials of interest \cite{simpson_magneto-optical_2016} or being unsuitable for magnetically shielded environments. 
Recent studies have, however, demonstrated AO magnetometry at low applied fields ($\sim$1\,mT) using NV ensembles with concentrations $\gtrsim$1\,ppm \cite{anishchik_low-field_2015, filimonenko_weak_2020, pellet-mary_relaxation_2023, dhungel_near_2024}.
In this low-field regime, NV-NV cross-relaxation from dipolar interactions, as well as NV spin-state mixing induced by local electric fields, can contribute to a magnetic-field-dependent reduction in AO-PL intensity, enabling sensitive AO magnetometry. 

In this work, we investigate room-temperature AO-PL from dense NV ensembles ($\sim$ppm) in CVD-grown diamond samples at low magnetic fields (<2\,mT). 
We experimentally characterize and numerically simulate AO-PL behavior as a function of the magnitude and direction of a weak applied magnetic field, finding reasonable agreement between measurements and calculations (Sec.~\ref{sec:full-spectrum}). 
The narrow linewidths of the observed low-field AO-PL features also allow us to identify NV hyperfine splittings for both $^{14}$N and $^{15}$N-enriched diamond samples. 
When NV hyperfine states are degenerate, they introduce additional depolarization channels for the spin populations, as observed in the relative AO-PL contrast. 
We further correlate the measured AO-PL dependence on applied magnetic field with nearly degenerate energy levels observed in microwave-based continuous-wave optically detected magnetic resonance (CW-ODMR) spectra. 
We then demonstrate AO DC magnetometry at about 1\,mT bias field, with comparable PL contrast and feature linewidths for AO-PL and CW-ODMR measurements (Sec.~\ref{sec:nuclear-hyperfine}). 
As a function of laser power, we experimentally find a maximum of AO-PL contrast. 
This effect is also captured by numerical simulations of a phenomenological model that includes both NV-concentration and degenerate-level dependent relaxation rates between NV spin sublevels, arising from NV-NV dipolar interactions (Sec.~\ref{sec:contrast_laser}).

\clearpage

\section{Background} \label{sec:background}

The negatively-charged NV center is a $C_{3v}$-symmetric point defect in diamond with ground and excited electronic spin triplet states ($S$ = 1). 
The electronic ground state Hamiltonian $H_{gs}$ for a single NV can be written as \cite{barry_sensitivity_2020}:
    \begin{equation} \label{eqn:gs}
        {H}_{gs}/{\hbar} = \vec{S} \cdot {D} \cdot \vec{S} + \gamma_{e} \vec{B} \cdot \vec{S} + \vec{S} \cdot {A} \cdot \vec{I} + \vec{I} \cdot {Q} \cdot \vec{I}.
    \end{equation}
Here, $\vec{S}$ and $\vec{I}$ are the electronic and nuclear spin operators, respectively, with $I=1$ for $^{14}$N and $I=1/2$ for $^{15}$N; 
${D}$ is the room-temperature zero field splitting (ZFS);
$\gamma_e$ is the electronic spin gyromagnetic ratio; $B$ is the applied magnetic field (nuclear Zeeman shifts are ignored due to negligible contribution at low magnetic fields);
${A}$ is the hyperfine interaction tensor;
and ${Q}$ describes the nuclear electric quadrupole coupling tensor for $^{14}$N (${Q} = 0$ for $^{15}$N).
For simplicity, we neglect the contributions from strain and electric fields.
The NV electronic ground ($^{3}A_{2})$ and excited ($^{3}E$) triplet states primarily follow spin-conserving optical transitions under 532\,nm laser irradiation, and emit broadband PL ($\approx$\,637–800\,nm). 
An alternative intersystem crossing to the singlet manifold preferentially allows excited NVs in the $\ket{m_s = \pm 1}$ spin states to decay to the ground $\ket{m_s = 0}$ state with reduced PL emission. As a result, optical excitation both induces spin-state-dependent PL and polarizes NV electronic spins to $\ket{m_s = 0}$ [Fig.~\ref{fig:Schematics}(a)] \cite{doherty_nitrogen-vacancy_2013}.

In an NV ensemble, NV-NV dipolar interactions increase the spin relaxation (depolarization) rate between the PL bright $\ket{m_s = 0}$ and dark $\ket{m_s = \pm 1}$ states \cite{choi_depolarization_2017}. 
The interaction Hamiltonian $H_{\text{int}}$ between two NV centers with electronic spin operators $\vec{S}_{1,2}$, and dipolar interaction strength $D_{\text{dd}}$ along a unit vector $\hat{n}_{12}$, is given by:
    \begin{equation} \label{eqn:int}
        {H}_{\text{int}} = D_{\text{dd}} [3(\vec{S}_{1} \cdot \hat{n}_{12})(\vec{S}_{2} \cdot \hat{n}_{12}) - (\vec{S}_{1} \cdot  \vec{S}_{2})].
    \end{equation} 
The ground state Hamiltonian for two interacting NVs is then \({H} = {H_{gs1}} + H_{gs2} + {H}_{\text{int}}\).
The depolarization rate is further enhanced when different NV centers have equal transition frequencies between electronic spin states. 
For such degenerate spin transitions, resonant NV-NV dipolar interactions induce spin state mixing, i.e., NV-NV cross-relaxation \cite{choi_depolarization_2017, kucsko_critical_2018}. 
Consequently, a reduction in both the NV spin polarization lifetime ($T_1$) and PL intensity can be observed [Fig.~\ref{fig:Schematics}(c)] \cite{pellet-mary_relaxation_2023}. 
Each NV within an ensemble in a single crystal diamond sample is oriented along one of the four crystallographic axes, with typically an equal fraction (one fourth) of the NV ensemble along each axis. 
By adjusting the applied magnetic field direction and magnitude in the low-field regime (<2\,mT), one can tune the spin transition frequencies of the four NV orientation classes and spectrally overlap their resonances. 
This magnetic field tuning enables careful measurement of low-field AO-PL signals, i.e., without applied microwaves. 

\begin{figure}
    \centerline{\includegraphics{Fig.1.pdf}}
    \caption{\label{fig:Schematics} 
    (a) NV energy levels and couplings allow optical initialization of electronic spin states and emission of spin-state-dependent photoluminescence (PL). 
    (b) AO-PL measurements utilize three sets of Helmholtz coils to control bias magnetic field magnitude and direction. NV centers in a single crystal diamond plate are optically excited by 532\,nm laser light and emit PL ($\approx$\,637–800\,nm), collected by a photodiode (not shown). A microwave antenna (not shown) enables comparison CW-ODMR measurements. Inset: dipolar interactions between different NV centers contribute to AO-PL contrast at low magnetic fields. 
    (c) Illustration of avoided crossing from two interacting near-resonant NV electronic spins at low magnetic fields (<2\,mT). Resonant NV-NV dipolar interactions at the avoided crossing increase the NV depolarization rate and reduce total PL emission.}
\end{figure}

\begin{table}[H]
\centering
\caption{\label{tab:samples} 
Diamond samples used in this study. All samples are electronic grade plates (few mm on each side and about 0.5\,mm thick), with a 10\,$\mu$m-thick surface layer of enhanced nitrogen and NV concentration ([N] = 16\,ppm, > 99.995\% $^{12}$C) as reported by Element Six Ltd. 
}
\begin{ruledtabular}
% \resizebox{\textwidth}{!}{%
\begin{tabular}{ccc}
\textbf{Sample \#}  & \textbf{{[}NV{]} (ppm)} & \textbf{N Isotope}\\
S1-14N                       & $\approx 3.8$                   & $^{14}$N \\
S2-15N                       & $\approx 3.8$                   & $^{15}$N \\
S3-14N                       & $\approx 3.8$                 & $^{14}$N \\
S4-14N                       & $\approx 2$                 & $^{14}$N \\
S5-14N                       & $\approx 0.3$                 & $^{14}$N \\
\end{tabular}%
% }
\end{ruledtabular}
\end{table}

Table~\ref{tab:samples} summarizes the characteristics of different CVD-grown NV-diamond samples employed in this work.
We observe NV-NV cross-relaxation features in low-field AO-PL measurements for diamond samples with NV concentration $\gtrsim$ 0.3\,ppm, corresponding to an average distance between NVs $\lesssim$ 26\,nm.

\begin{figure*}[hbt!]
    \centering
    \includegraphics{Fig.2.pdf}
    \caption{\label{fig:concentration_spectrum} 
    (a) Illustration of applied magnetic fields and four NV orientations along unit vectors $\hat{n}_\lambda$, $\hat{n}_\phi$, $\hat{n}_\chi$ and $\hat{n}_\kappa$. The on-axis magnetic field $B_{\parallel}$ is along the [111] crystallographic axis. The off-axis magnetic field $B_{\perp}$ is along the [$\bar{1}$10] crystallographic axis. $\theta$ is the polar angle from $B_{\parallel}$.
    (b) Experimentally determined AO-PL contrast as a function of applied magnetic fields using sample S1-14N with $\approx 3.8$\,ppm NV concentration. Large AO-PL contrast around zero applied field, as well as line features at specific $\theta$ values are observed.
    (c) Simulated AO-PL contrast using a fixed dipolar interaction strength.
    (d) Expanded view of the upper right quadrant in Fig.~\ref{fig:concentration_spectrum}(c), with labels at specific line features indicating the cross-relaxation between NVs of orientations along 
        (i) $\hat{n}_\phi$, $\hat{n}_\chi$ and $\hat{n}_\kappa$ at $\theta = 0$\textdegree;
        (ii) $\hat{n}_\kappa$ at $\theta = 22.2$\textdegree;
        (iii) $\hat{n}_\lambda$ and $\hat{n}_\chi$; $\hat{n}_\phi$ and $\hat{n}_\kappa$ at $\theta = 39.3$\textdegree;
        (iv) $\hat{n}_\lambda$ and $\hat{n}_\kappa$ at $\theta = 58.5$\textdegree;
        (v) $\hat{n}_\lambda$ and $\hat{n}_\phi$ at $\theta = 90$\textdegree.
    NV hyperfine interactions contribute to parallel line structures within AO-PL contrast features.
    }
\end{figure*}
% \noindent
\section{Results}
\subsection{NV-NV Cross-relaxation Features for a Dense Ensemble}  \label{sec:full-spectrum}

\begin{figure*}[!htb]
    \centering
    \includegraphics[width=\textwidth]{Fig.3.pdf}
    \caption{\label{fig:14vs15}
    (a) Experimentally measured AO-PL contrast as a function of $B_{\parallel}$ and $B_{\perp}$ using $^{14}$N-enriched sample S1-14N (top) and $^{15}$N-enriched sample S2-15N (bottom) with the same NV concentration ($\approx$ 3.8\,ppm). Horizontal dashed lines (orange for S1-14N and purple for S2-15N) indicate line-cut of data shown in Fig.~\ref{fig:14vs15}(b) for $B_{\parallel}$ = 1.24\,mT. 
    (b) Experimentally measured AO-PL contrast for samples S1-14N and S2-15N (solid lines) and numerical simulations (dashed lines) at fixed $B_{\parallel} = $ 1.24\,mT. Vertical offsets are applied to the simulation results for visual clarity. In each measurement as $B_{\perp}$ approaches 0\,mT, there are overlapping, unresolved AO-PL contrast peaks from the increased number of near-degenerate spin transitions for all NV orientations. 
    At $B_{\perp} \approx 0.5$\,mT, the limited SNR in experimental measurements hinders the identification of cross-relaxation features predicted by numerical simulations.
    At $B_{\perp} \approx 1.05$\,mT, there are five (three) AO-PL contrast peaks with separation $\Delta B_{14N} \approx 0.09$\,mT ($\Delta B_{15N} \approx 0.13$\,mT) and amplitude ratio of 1:2:3:2:1 (1:2:1) in the top (bottom) data and simulations, respectively, consistent with the effect of NV hyperfine interactions in the two samples.
}
\end{figure*}
\noindent

Fig.~\ref{fig:Schematics}(b) depicts the experimental setup used to study low-field AO-PL, including NV-NV cross-relaxation features. 
An NV diamond sample is exposed to a controllable magnetic field, with magnitude and direction determined by three sets of Helmholtz coils aligned along the diamond's [110], [$\bar{1}$10], and [001] crystallographic axes. 
The unit vectors parallel to the NV symmetry axes are $\hat{n}_\lambda$, $\hat{n}_\phi$, $\hat{n}_\chi$, and $\hat{n}_\kappa$ along $[111]$, $[\bar{1}\bar{1}1]$, $[1\bar{1}\bar{1}]$, and $[\bar{1}1\bar{1}]$, respectively. 
As illustrated in Fig.~\ref{fig:concentration_spectrum}(a), the on-axis magnetic field $B_{\parallel}$ is defined as parallel to NVs along axis $\hat{n}_\lambda$; and the off-axis magnetic field $B_{\perp}$ is perpendicular to both axes $\hat{n}_\lambda$ and $\hat{n}_\phi$. 
The AO-PL contrast for each on/off-axis magnetic field value ($B_{\parallel}$, $B_{\perp}$) is determined by the normalized difference in intensities between a measured AO-PL signal $I_\text{sig}$ and an AO-PL reference $I_\text{ref}$. $I_\text{ref}$ is given by the maximum measured AO-PL from a given scan of ($B_{\parallel}$, $B_{\perp}$) values in which the NV orientation classes are spectrally separated. The AO-PL contrast is then calculated as \(C = 1 - I_{\text{sig}} / I_{\text{ref}}\). Further details of the experimental procedure and calibration are given in Appendix~\ref{supp:exp}.

We experimentally determine the low-field AO-PL contrast as a function of $B_{\parallel}$ and $B_{\perp}$, starting with a 3.8\,ppm NV concentration diamond enriched with $^{14}$N (sample S1-14N) [Fig.~\ref{fig:concentration_spectrum}(b)]. 
We vary both $B_{\parallel}$ and $B_{\perp}$ over a range of about 1.5 to $-1.5$\,mT, with step size $\approx$\,0.02\,mT. 
The resulting two-dimensional plot of experimental AO-PL contrast is in reasonable agreement with a numerical simulation based on the simple two-NV coupled Hamiltonian described by Eqs.~(\ref{eqn:gs},~\ref{eqn:int}) with a fixed dipolar interaction strength [Fig.~\ref{fig:concentration_spectrum}(c)] \cite{anishchik_method_2019}. 
Differences between the experimental and simulated magnitudes of AO-PL contrast line features likely arise from the simplicity of the model used.
Details of the numerical simulation are included in Appendix~\ref{supp:Jeyson_model}. 
When the applied magnetic field is near zero ($\lesssim$\,0.1\,mT), spin transitions for NVs along all four axes are spectrally overlapped, resulting in maximum NV-NV cross-relaxation and AO-PL contrast. 
The relatively large magnetic field linewidth measured in AO-PL contrast features at near-zero-field may be attributed to the influence of local electric fields \cite{mittiga_imaging_2018, pellet-mary_relaxation_2023} and/or magnetic field and strain gradients.
The experimental AO-PL contrast maximizes at $B_\parallel \approx-0.02$\,mT, likely due to a background magnetic field (e.g., from the Earth and other lab instrumentation). 
Additionally, significant AO-PL contrast is observed in both experiment and simulation at specific polar angles $\theta$ from $B_{\parallel}$, consistent with the effect of cross-relaxation among different NV orientations resulting from near-degenerate spin transitions \cite{filimonenko_weak_2020, pellet-mary_relaxation_2023, dhungel_near_2024}. 
For example, at $\theta$ = 0\textdegree\ ($B_{\parallel} \neq 0$ and $B_{\perp}$ = 0), the magnetic field is aligned with $\hat{n}_\lambda$ and the three non-aligned NV orientations (along $\hat{n}_\chi$, $\hat{n}_\phi$ and $\hat{n}_\kappa$) experience equal magnetic field projections. 
In the experiment, this configuration produces a broad, stripe-like feature with large AO-PL contrast (vertical around $B_{\perp}$ = 0 in Fig.~\ref{fig:concentration_spectrum}(b)) due to the enhanced cross-relaxation between the three overlapping NV spin resonances. 
The simulation using the simple two-NV model does not fully capture this broad stripe feature, as seen in Fig.~\ref{fig:concentration_spectrum}(c, d).
This discrepancy may arise from ensemble averaging in the experiment, including magnetic field and strain gradients, as well as variation in NV spacing.

At $\theta$ = 39.3\textdegree, two sets of NV orientations (along $\hat{n}_\lambda$ and $\hat{n}_\chi$; $\hat{n}_\phi$ and $\hat{n}_\kappa$) experience the same applied field magnitude, with resulting degeneracies in NV spin transitions. 
Here, cross-relaxation from the two sets of degenerate NV orientations creates a line feature with moderate AO-PL contrast, in both experiment and simulation, compared to the results at $\theta$ = 0\textdegree\ with three degenerate NV orientations. 
Additional line features of moderate AO-PL contrast are observed in both experiment and simulation for other specific polar angles that induce degeneracies between pairs of NV orientations, as seen in Fig.~\ref{fig:concentration_spectrum}(b, c, d).
At $\theta$ = 22.2\textdegree, where the total applied field is transverse to NVs oriented along $\hat{n}_\kappa$, there is a line feature with very weak AO-PL contrast. 
A comprehensive analysis of all AO-PL contrast features given by the simulation is presented in Appendix~\ref{supp:Jeyson_analysis}, including for different classes of interactions (Fig.~\ref{sup-fig:decomposition}) and for specifically oriented interacting NV center pairs (Fig.~\ref{fig:decomposition-planes}). 
Additional measurements on a $^{14}$N-enriched diamond sample with 0.3\,ppm NV concentration (sample S5-14N) are presented in Appendix~\ref{supp:spectra-concentration}, with Fig.~\ref{sup-fig:concentrationComparison} showing cross-relaxation features at the same angles in the ($B_{\parallel}$, $B_{\perp}$) plane as for sample S1-14N, but with much smaller AO-PL contrast due to the weaker dipolar coupling between lower concentration NVs in sample S5-14N.

\subsection{Hyperfine Interaction and NV-NV Cross-relaxation} \label{sec:nuclear-hyperfine}
\begin{figure*}[!htb]
    \centerline{\includegraphics{Fig.4.pdf}}
    \caption{\label{fig:15N_lockin} 
    (a) Normalized AO-PL lock-in amplifier (LIA) measurements as a function of $B_{\perp}$ from $^{15}$N-enriched sample S2-15N at fixed $B_\parallel = $ 1.24\,mT. Focus is on LIA features around 0.5\,mT (purple) and 1\,mT (orange).
    (b) AO-PL LIA signals with a smaller $B_\perp$ step size ($\approx$ 0.002\,mT) from shaded areas in (a). Two weak features are observed in the shoulder of the dispersive signal with central zero crossing at about 0.53\,mT (left). Near 1\,mT, three dispersive features are measured with splitting $\Delta B_{15N} \approx 0.13$\,mT (right). For both (left) and (right), vertical dashed lines indicate near-degeneracy of hyperfine transitions.
    (c) Summary of measurements of microwave-based CW-ODMR NV spin transition frequencies as a function of $B_{\perp}$, including different NV orientations and hyperfine splitting, for the same sample (S2-15N) and at fixed $B_\parallel = $ 1.24\,mT. 
    Between $B_{\perp} \approx$ 0.4\,mT and 0.65\,mT (left), NVs along orientation $\hat{n}_\kappa$ experience near-zero total applied magnetic field, leading to multiple avoided crossings between the hyperfine-split $\ket{m_s = +1 \leftrightarrow 0}$ and $\ket{m_s = -1 \leftrightarrow 0}$ spin transitions. 
    Between $B_{\perp} \approx$ 0.8\,mT and 1.3\,mT (right), results are shown for all NV orientations, including only hyperfine-split $\ket{m_s = +1 \leftrightarrow 0}$ transitions for better clarity. (Consistent CW-ODMR results are found for $\ket{m_s = -1 \leftrightarrow 0}$ transitions.) 
    Two groups of NV spin resonances are observed, each of which consists of two different NV orientations (along $\hat{n}_\lambda$ and $\hat{n}_\chi$; $\hat{n}_\phi$ and $\hat{n}_\kappa$). 
    Within each spin resonance group, a single hyperfine resonance overlaps at both $B_{\perp} \approx$ 0.93\,mT and 1.19\,mT; whereas two hyperfine resonances overlap at $B_\perp \approx$ 1.05\,mT. $B_\perp$ values for near-degenerate spin transition frequencies align well with those of zero crossings in the AO-PL LIA signals in (b), indicated by vertical dashed lines.
    }
\end{figure*}

The experimental and simulation results in Fig.~\ref{fig:concentration_spectrum} also display multiple parallel structures within each $\theta$-dependent AO-PL contrast feature. 
To investigate the role of the NV hyperfine interaction in this AO-PL sub-structure, we conduct comparative measurements on $^{14}$N and $^{15}$N samples with the same NV concentration (samples S1-14N and S2-15N). 
Fig.~\ref{fig:14vs15}(a) shows AO-PL experimental results for positive-only values of $B_{\parallel}$ and $B_{\perp}$, taken with a smaller magnetic field step size ($\approx$ 0.01\,mT) than the results in Fig.~\ref{fig:concentration_spectrum}(b). (Similar results are found for the other ($B_{\parallel}$, $B_{\perp}$) quadrants.)  
In addition, we measure AO-PL contrast at fixed $B_{\parallel}$ = 1.24\,mT by scanning $B_{\perp}$ with an even finer step size ($\approx$ 0.002\,mT) to resolve the individual cross-relaxation line shapes more clearly (solid lines in Fig.~\ref{fig:14vs15}(b)). 
As $B_{\perp}$ approaches 0\,mT, there are overlapping, unresolved AO-PL contrast peaks due to the increasing number of spin transition degeneracies between different NV orientations. 
At $B_{\perp}\approx 1.05$\,mT, equivalent to $\theta$ $\approx$ 40\textdegree\ in Fig.~\ref{fig:14vs15}(a), two sets of NV orientations (along $\hat{n}_\lambda$ and $\hat{n}_\chi$; $\hat{n}_\phi$ and $\hat{n}_\kappa$) are near-degenerate. 
In this parameter regime, the NV hyperfine interaction provides additional depolarization channels and splits the NV-NV cross-relaxation resonances, with associated AO-PL contrast features. 
In particular, the number of AO-PL contrast peaks varies between the two samples. The $^{14}$N spectrum in Fig.~\ref{fig:14vs15}(b, top) shows five peaks with a separation $\Delta B_{14N} \approx 0.09$\,mT between neighboring peaks. 
The ratio of individual peak amplitudes follows 1:2:3:2:1, as expected for the $^{14}$N nuclear spin quantum number $I = 1$.  
Similarly, the $^{15}$N spectrum in Fig.~\ref{fig:14vs15}(b, bottom) displays three peaks with $\Delta B_{15N} \approx 0.13$\,mT, and a ratio of individual peak amplitude given by 1:2:1, consistent with $I=1/2$ for $^{15}$N. 
The magnetic field separation between neighboring peaks can be calculated from the nitrogen nuclear hyperfine splitting $A_N$ and the projection angles of the NVs as \(\Delta B_N = A_N / [\gamma_{e}\text{cos}(\alpha)]\), where $\gamma_{e}$ is the gyromagnetic ratio of the NV electronic spin and $\alpha$ is the angle between $B_{\perp}$ and NVs along axis $\hat{n}_{\chi}$ or $\hat{n}_{\kappa}$ (see Appendix~\ref{supp:hyperfine_separation}).

These experimental results are in good agreement with calculations using the same model from Sec.~\ref{sec:full-spectrum} (dashed lines in Fig.~\ref{fig:14vs15}(b)). 
However, the limited experimental signal-to-noise ratio (SNR) hinders identification and analysis of AO-PL features with modest contrast (e.g., near $B_\perp \approx 0.5$\,mT). 
To address this issue, we perform lock-in measurements of AO-PL from the $^{15}$N sample (S2-15N), with its simpler hyperfine structure, by amplitude modulating $B_\perp$. Fig.~\ref{fig:15N_lockin}(a) shows the normalized in-phase output (X) of the lock-in amplifier (LIA) as a function of $B_\perp$, for fixed $B_{\parallel} = 1.24$\,mT. This LIA measurement is effectively a higher SNR derivative of the AO-PL signal of Fig.~\ref{fig:14vs15}(b, bottom). 
We highlight two regions of the LIA results that resolve distinct dispersive AO-PL features from NV-NV cross-relaxation: small ``shoulders'' for $B_\perp$ near 0.46\,mT and 0.6\,mT in the wings of a prominent feature at $B_\perp \approx 0.53$\,mT [Fig.~\ref{fig:15N_lockin}(b, left)]; and a triplet feature split by $\Delta B_{15N} \approx 0.13$\,mT around $B_\perp \approx 1.05$\,mT [Fig.~\ref{fig:15N_lockin}(b, right)], matching well the results in Fig.~\ref{fig:14vs15}(b).

To further characterize the effects of NV-NV cross-relaxation and the NV hyperfine interaction on PL features, we perform microwave-based CW-ODMR measurements on sample S2-15N as a function of $B_\perp$ and for fixed $B_\parallel = 1.24$\,mT, with results summarized in Fig.~\ref{fig:15N_lockin}(c). 
These microwave-based measurements extract the spin transition frequencies for all NV orientations and determine the values of $B_{\perp}$ where these spin transitions become nearly degenerate. 
In addition, CW-ODMR measurements clearly resolve small (few MHz) splittings from NV hyperfine interactions. 
In the range of $B_\perp \approx$ 0.4\,mT to 0.65\,mT, NVs of the same orientation (along $\hat{n}_\kappa$) exhibit several nearly degenerate (avoided crossing) spin transitions in the CW-ODMR spectrum [Fig.~\ref{fig:15N_lockin}(c, left)], which correspond well with observed features in AO-PL LIA measurements [Fig.~\ref{fig:15N_lockin}(b, left)]. 
For $B_\perp \approx 1.05$\,mT, CW-ODMR measurements show near-degeneracies in both hyperfine resonances from two sets of NV orientations (along $\hat{n}_\lambda$ and $\hat{n}_\chi$, or $\hat{n}_\phi$ and $\hat{n}_\kappa$); whereas for the two nearby shoulder features (at $B_\perp \approx$ 0.92\,mT and 1.18\,mT), only one hyperfine resonance has a near-degeneracy for each set of NV orientations [Fig.~\ref{fig:15N_lockin}(c, right)]. 
These CW-ODMR spectroscopic measurements are consistent with the $B_\perp$ values and amplitude ratios for the observed AO-PL LIA features, see dashed vertical lines in Fig.~\ref{fig:15N_lockin}; and also with the ratio of AO-PL contrast peak amplitudes for sample S2-15N shown in Fig.~\ref{fig:14vs15}(b, lower right). 
Comparing to the AO-PL LIA measurements for $B_\perp \approx$ 0.4\,mT and 0.65\,mT, the increased number of NVs with near-degenerate spin transitions at $B_\perp \approx$ 0.92\,mT, 1.05\,mT, and 1.18\,mT contributes to the larger AO-PL LIA output amplitudes at the larger $B_\perp$ values [Fig.~\ref{fig:15N_lockin}(a)]. 

We also demonstrate DC magnetometry using AO-PL LIA measurements near the center of the triplet feature in sample S2-15N, with $B_\perp \approx 1.05$\,mT and fixed $B_\parallel = 1.24$\,mT. We determine a DC sensitivity $\approx 30\,{\rm nT/\sqrt{Hz}}$ (see Appendix~\ref{supp:sensitivity} and Fig.~\ref{sup-fig:lockin_sensitivity}). The measured AO-PL feature contrast and linewidth are comparable to those obtained with microwave-based CW-ODMR for similar NV-diamond samples and experimental conditions (see Appendices~\ref{supp:exp}, \ref{supp:hyperfine_separation}, and \ref{supp:sensitivity}, and Figs.~\ref{sup-fig:fieldCalibration}, \ref{sup-fig:15N_peakSeparation}, and \ref{sup-fig:lockin_sensitivity}).  These results suggest that the DC magnetic sensitivity of low-field AO NV measurements can approach that of conventional CW-ODMR, but without the experimental complexity of microwaves \cite{barry_sensitivity_2020}.

\subsection{Dependence of AO-PL Contrast on Laser Power} \label{sec:contrast_laser}

\begin{figure}
    \centerline{\includegraphics{Fig.5.pdf}}
    \caption{\label{fig:laser_dependency} AO-PL contrast at $B_{\parallel} = 1.24$\,mT and $B_\perp = 1.05$\,mT as a function of laser power and intensity for samples S3-14N ([NV] $\approx$ 3.8\,ppm) and S4-14N ([NV] $\approx$ 2\,ppm). The laser intensity is calculated by approximating a Gaussian excitation laser beam profile of radius $\approx$ 30\,$\mu m$ (see Appendix~\ref{supp:exp}). Markers indicate values determined from experimental measurements, with the standard deviation given by error bars; solid lines are from a rate-equation model with NV-concentration and spin-resonance-dependent relaxation rates between spin sublevels. For both samples, the contrast exhibits a maximum determined by the trade-off between optical pumping and spin relaxation.  Sample S3-14N has higher overall AO-PL contrast, for all laser powers, because of its larger [NV] and hence stronger NV-NV dipolar interactions and cross-relaxation features.}
\end{figure}

We next characterize AO-PL contrast as a function of laser power for two $^{14}$N samples at fixed $B_{\parallel} = 1.24$\,mT and $B_\perp = 1.05$\,mT, i.e., near the maximum contrast from the quintet of hyperfine peaks, as in Fig.~\ref{fig:14vs15}(b, top).
The reference AO-PL contrast is measured at $B_\perp = 0.73$\,mT where all NV orientation classes are spectrally separated.
The two samples are from the same growth process, with identical nitrogen concentration ([N] $\approx$ 16\,ppm) but different electron irradiation doses, which allows varying NV concentration while keeping other material properties and experimental conditions (laser spot size, etc.) constant: sample S3-14N ([NV] $\approx$ 3.8\,ppm) and sample S4-14N ([NV] $\approx$ 2\,ppm), see Table~\ref{tab:samples}. 
Fig.~\ref{fig:laser_dependency} shows the measured change of AO-PL contrast for each sample, with laser power varied from 0.1\,mW to 50\,mW, in reasonable agreement with numerical results from a rate-equation model.
In this model, we include relaxation between $\ket{m_s = \pm 1}$ and $\ket{m_s = 0}$ spin sublevels in both the electronic ground and excited states. 
Details of the rate-equation model, supporting measurements, and results for hyperfine quintet peak widths as a function of laser power are given in Appendix~\ref{supp:AOPL_power}.

For both samples, the observed AO-PL contrast initially increases with laser power, reaches a maximum value, and then decreases at higher power. 
We attribute the initial increase of AO-PL contrast to improved spin polarization from optical pumping, consistent with the low-field AO-PL measurements from Ref. \cite{dhungel_near-zero-field_2024}. 
At higher laser power, optical pumping polarizes NV centers to the $\ket{m_s = 0}$ state at a rate faster than relaxation between the bright $\ket{m_s = 0}$ and dark $\ket{m_s = \pm 1}$ states, decreasing AO-PL contrast. 
Including the decay rates between spin sublevels to be both NV-concentration and spin-resonance-dependent \cite{choi_depolarization_2017, kucsko_critical_2018}, our rate-equation model reproduces the experimentally observed shift of maximum AO-PL contrast to higher laser power for sample S3-14N, with its larger [NV] and hence stronger NV-NV dipolar interactions; see solid lines in Fig.~\ref{fig:laser_dependency}. 
Note that at high 532\,nm laser excitation intensities, NV ionization and charge cycling may occur, decreasing the population of negatively-charged NVs via conversion to the neutral charge state (NV$^0$), and thereby further reducing AO-PL contrast \cite{edmonds_characterisation_2021, giri_selective_2019, cardoso_barbosa_impact_2023}.

\section{Conclusion} \label{sec:discussion}
In summary, we experimentally investigate all-optical (AO), microwave-free photoluminescence (PL) in NV ensembles in diamond as a function of the magnitude and direction of low magnetic fields (<2\,mT).  We observe increases in AO-PL contrast arising from near-degeneracies in NV spin transitions, with contributions from different NV orientations and hyperfine splitting in both $^{14}$N ($I = 1$) and $^{15}$N ($I = 1/2$) diamonds.
Measurements are in reasonable agreement with results from numerical simulations using a two-NV model of dipolar interactions. 
Further substantiation of this physical picture is provided by consistency between the magnetic field values of measured AO-PL contrast peaks and those for near-degenerate NV spin and hyperfine transitions observed in CW-ODMR spectra.
Experimentally, we also find a maximum of AO-PL contrast as a function of applied laser power, using two $^{14}$N diamond samples of different NV concentration. 
These measurements are consistent with numerical results based on a model with NV-concentration and spin-resonance-dependent relaxation rates between both ground and excited state $\ket{m_s = \pm 1}$ and $\ket{m_s = 0}$ spin sublevels. 

Our results indicate that an NV spin optical pumping rate comparable to the depolarization rate from NV-NV cross-relaxation yields optimal low-field AO-PL contrast. 
By using a diamond sample with higher NV concentration, which increases dipolar interactions and associated spin relaxation, one may be able to use higher laser excitation power to achieve improvements in both AO-PL contrast and total photon emission, which can be expected to provide improved AO NV magnetic field sensitivity. 
Future work in this direction may include experimental study and modeling of the role of NV ionization and charge-state dynamics at higher laser power. 
Although AO-PL contrast is largest at near-zero-field, as NVs from all four orientations in the diamond host can resonantly interact, we find the magnetic field linewidth of near-zero-field AO-PL features to be $\approx 2.5\times$ broader than at finite fields $\approx$\,1\,mT (see Appendix~\ref{supp:slope_comparison}). The increased feature linewidth at near-zero-field may arise from NV spin-state mixing induced by local electric fields \cite{mittiga_imaging_2018, pellet-mary_relaxation_2023} and/or magnetic field and strain gradients in the system. 
Therefore, the optimal conditions for near-zero-field AO NV magnetic sensing may rely on both diamond engineering and technical optimization.

The AO, microwave-free approach to ensemble NV magnetometry offers opportunities to reduce device complexity and avoid introducing disturbances to sensing targets \cite{vorst_rf_2006, schlussel_wide-field_2018, zhang_battery_2021}. 
In particular, given the robustness of the diamond host and NV properties \cite{barry_sensitivity_2020}, low-field AO measurements may be performed in harsh environments (extreme temperature, pressure, radiation, etc.), where efficient microwave delivery and application of a substantial bias magnetic field are challenging or may adversely affect the system under study \cite{fu_sensitive_2020}.
With optimal AO-PL contrast occurring at modest optical pumping power (mW-scale), AO operation may allow sensitive NV magnetometry in a device of relatively low size, weight, and power (low-SWaP), enabling diverse applications beyond a controlled laboratory environment. 
Finally, our studies may prove advantageous when extending AO measurements to other defect ensemble systems with spin-dependent optical properties, such as in silicon carbide (SiC) \cite{widmann_coherent_2015, simin_all-optical_2016} and hexagonal boron nitride (h-BN) \cite{gottscholl_initialization_2020, gottscholl_room_2021}.

\section*{Acknowledgments}
We thank Jner Tzern Oon for helpful discussions on implementing NV-NV dipolar interactions in numerical simulations; and Jiashen Tang for assistance in performing polarization lifetime measurements. 
We acknowledge the U.S. Army Research Laboratory for providing NV diamond samples.
J.\,T.\,-I. acknowledges support from the Universidad Nacional de Colombia, project No. 57522 "Centro de Excelencia en tecnologías cuánticas y sus aplicaciones a metrología" Código: 57522.
M.\,C acknowledges partial funding from NSF-DMR-2226956.
This work is supported by, or in part by, the Department of Energy under Grant. No. DESC0021654; U.S. Army Research Laboratory under Contract No. W911NF2420143; and the University of Maryland Quantum Technology Center.

\section*{Data Availability}
The data that support the findings of this paper are openly available \cite{DVN/6TFFPG_2025}.

\appendix
\section{Experimental Details} 
\label{supp:exp}
\subsection{Experiment Methods}
532\,nm CW laser light (Lighthouse Photonics Sprout-H-10W) is focused into a diamond sample with a microscope objective (Olympus 20X/NA 0.45) to optically excite the NV centers, with a typical Gaussian excitation laser beam profile of radius $\approx$ 30\,$\mu m$. NV-diamond sample properties are summarized in Table~\ref{tab:samples} of the main text. A half-wave plate (Thorlabs WPHSM05-532) is mounted before the microscope objective to control the optical polarization direction. NV photoluminescence (PL) is collected through the same objective, directed with a dichroic mirror (Semrock FF552-DI02), filtered with a long-pass filter (Semrock BLP01-532R-25), and imaged onto an avalanche photodiode (Thorlabs APD410A) through a 50\,mm lens (Thorlabs LA1131-B-ML). A polarizing beamsplitter (Thorlabs PBS12-532-HP) and a half-wave plate (Thorlabs WPHSM05-532) held within a rotation mount (Thorlabs CRM05) are used to control the laser power delivered to the diamond sample. The main text provides details of diamonds used in this study, including NV ensemble properties in each sample.

A bias magnetic field of controllable magnitude and direction is provided by three-axis Helmholtz coils (Ferronato BH300-3-A) connected to a three-channel power supply (Rohde \& Schwarz HMP4030). Relay boards are used to alter the magnetic field directions. The diamond plate under study is attached to a dielectric mirror (Thorlabs BB1-E02) to boost optical collection efficiency. The diamond-mirror system is then mounted on a piezoelectric rotational stage (Thorlabs ELL14) to align the longitudinal and transverse directions of the [111] NV-axis to the applied magnetic field. An additional coil connected to a microwave signal generator (SRS SG384) is used to perform microwave-based continuous-wave optically detected magnetic resonance (CW-ODMR) measurements.

The output of the photodiode used to collect NV PL is sent to a data-acquisition device (National Instruments USB-6363) for direct measurements, or to a lock-in amplifier (SRS SR865) for lock-in detection of all-optical (AO) PL. To apply small magnetic field amplitude modulation as part of lock-in AO-PL measurements, the output of a lock-in amplifier (LIA) local frequency generator is connected to the Helmholtz coils through a Class-D audio amplifier (Sure Electronics TDA7492).

\subsection{Calibration of Applied Magnetic Fields} 
\label{supp:field-calib}
\begin{figure*}[!htb]
    \centering
    \includegraphics[width=0.8\textwidth]{Fig.6.pdf}
    \caption{\label{sup-fig:fieldCalibration}
    (a) Schematic of unit vectors parallel to the four NV symmetry axes within a diamond sample, relative to the bias magnetic field Cartesian coordinates, with $\hat{z}$ vertical in the lab frame.
    (b) CW-ODMR spectra measured from sample S2-15N with 0.9\,A, 1.8\,A, and 2.7\,A applied to the y-axis Helmholtz coil, with the diamond oriented such that the coil generates a magnetic field only along the [100] crystallographic orientation. Vertical offsets are applied to the spectra for better clarity. The results from x and z-axis coils behave similarly.
    (c) Lorentzian fit to measured CW-ODMR spectra yields magnetic field along y-axis as a function of current applied to the y-axis Helmholtz coil. Linear fit (solid line) produces a current-to-magnetic-field calibration constant $k_y$ = 0.4068 $\pm$ 0.0009\,mT/A. 
}
\end{figure*}
\noindent

We calibrate magnetic fields generated by the three-axis Helmholtz coils by first rotating the diamond sample so that the magnetic field in $\hat{y}$ produced by one set of coils is along the diamond's [100] crystallographic orientation. As a result, the magnetic fields from coils along the x, y, and z axes individually project with equal magnitude onto all four NV orientations within a single-crystal diamond sample, illustrated in Fig.~\ref{sup-fig:fieldCalibration}(a). We then perform CW-ODMR measurements by using a microwave coil connected to a microwave signal source to coherently manipulate the NV ground state electronic spin states. The central hyperfine resonance frequencies from both the $\ket{m_s = +1}$ to $\ket{m_s = 0}$ transition and $\ket{m_s = -1}$ to $\ket{m_s = 0}$ transition can be extracted through a multi-Lorentzian fit to CW-ODMR data, with good overlap of the CW-ODMR spectra from all four NV orientations. Next, we calculate the applied magnetic field amplitude \(B = \Delta f / [2\gamma_e\cos(109.47^{\circ}/2)]\), where $\Delta f$ is the frequency difference between the $\ket{m_s = \pm 1}$ to $\ket{m_s = 0}$ transitions, $\gamma_e$ is the gyromagnetic ratio of the NV electronic spin, and $\cos(109.47^{\circ}/2)$ is the angle factor due to the applied magnetic field being aligned along the diamond's [100] orientation and not along the [111] symmetry axis. By repeating this procedure for the fields produced from each of the three Helmholtz coils, and performing linear fitting of the calculated magnetic fields to the corresponding applied currents, we obtain the current-to-magnetic-field calibration constants for each of the three directions. Fig.~\ref{sup-fig:fieldCalibration}(b, c) show example measured CW-ODMR spectra for different applied currents to the y-axis Helmholtz coil, and a linear fit to extract the calibration constant in $\hat{y}$. Using this method, we obtain the current-to-magnetic-field calibration constants $k_x$ = 0.406 $\pm$ 0.002\,mT/A, $k_y$ = 0.4068 $\pm$ 0.0009\,mT/A, and $k_z$ = 0.5119 $\pm$ 0.0008\,mT/A for the three directions in the lab frame.

To apply magnetic fields in the configuration described in the main text, with on-axis field $B_\parallel$ along [111] and off-axis field $B_\perp$ along [$\overline{1}$10] crystallographic axes, we rotate the diamond sample around the z axis by 45\textdegree. $B_\parallel$ is then described by a linear combination of $B_y$ and $B_z$, with $B_z = B_y / \sqrt{2}$; and $B_\perp$ is given by $B_x = I_x k_x$ in the lab frame. For example, setting $I_y = 2$\,A for the y-axis coil yields $B_y = I_y k_y = 0.814\,\text{mT}$ and hence $B_z = B_y / \sqrt{2} = 0.575\,\text{mT}$, equivalent to applying $I_z = B_z / k_z = 1.12$\,A. In this example, $B_\parallel = \sqrt{B_y^2 + B_z^2} = 0.997\,\text{mT}$.
The alignment of applied magnetic fields is further characterized by comparing the polar angles $\theta$ of different observed AO-PL cross-relaxation features with the expected values from symmetry considerations. Our findings indicate that the experimental alignment is consistent with theoretical expectations, with magnetic field misalignment limited to less than 1\textdegree.

\subsection{Dependence of Lock-in Output Amplitude on Magnetic Field Modulation Frequency} 
\label{supp:lockin-freq}
\begin{figure}[!htb]
    \centering
    \includegraphics[width=0.45\textwidth]{Fig.7.pdf}
    \caption{\label{sup-fig:ModulationFreq}
    Experimental demodulated AO-PL LIA spectrum from NV-NV cross-relaxation for sample S3-14N around $B_\perp \approx 1.05$\,mT and at fixed $B_\parallel = 1.24$\,mT for modulation frequency $f_m$ of 100\,Hz (blue) and 500\,Hz (orange). The upper trace is multiplied by 5 and shifted upwards for better clarity. A significant phase difference can be seen between the two traces, along with about 14 times lower peak amplitude for the higher modulation frequency.
}
\end{figure}
\noindent

We observe a dependence of the LIA output amplitude on the modulation frequency $f_m$ as we amplitude modulate $B_{\perp}$. Fig.~\ref{sup-fig:ModulationFreq} presents an example measured demodulated AO-PL LIA spectrum around $B_\perp \approx 1.05$\,mT and at fixed $B_\perp = 1.24$\,mT for sample S3-14N. The phase from the LIA is set to 0\textdegree\ so that the amplitude measured with $f_m = 100$\,Hz is maximized. 
Although higher $f_m$ is desired to avoid low frequency noise and improve signal-to-noise ratio (SNR), we find the peak amplitude from the central demodulated spectral feature decreases by a factor of about 14 as $f_m$ increases from 100\,Hz to 500\,Hz. In addition, there is a noticeable distortion (i.e., phase shift) of the dispersive line shapes at higher $f_m$. These observations are similar to the results reported in Ref.~\cite{anishchik_sensitive_2017} where adiabatic exchange of NV electronic spin population across the avoided crossings at low $f_m$ produces larger demodulated output amplitude. In order to resolve various AO-PL line shapes from NV-NV cross-relaxation more clearly without compromising the SNR from low frequency noise, we perform all LIA measurements reported in the main text with $f_m = 100$\,Hz. The larger demodulated peak amplitude at this modulation frequency helps to resolve the two shoulder signals from dipolar interactions between NVs along $\hat{n}_{\kappa}$ in Fig.~\ref{fig:15N_lockin}(c, left) in the main text.

\subsection{Calibration of Magnetic Field Amplitude Modulation Strength} 
\label{supp:mod-calib}

As described in the previous section, we perform AO-PL LIA measurements by amplitude modulating $B_\perp$ at a rate (modulation frequency $f_m$) of 100\,Hz. The SRS lock-in amplifier (LIA) outputs a sine wave with 0.1$V_{pp}$, which is passed through an audio amplifier to drive current through the x-axis Helmholtz coil and generate the modulated field. A 3.5\,$\Omega$ resistor is connected in series to the Helmholtz coil to improve impedance matching and provide stronger modulation amplitude. To calibrate the $B_\perp$ modulation amplitude produced by the Helmholtz coil, we perform lock-in CW-ODMR measurements. We apply an unmodulated bias field in the [$\overline{1}$10] direction so that the measurement can be optimally sensitive to $B_\perp$ modulation in the same direction. The microwave signal generator used for CW-ODMR outputs frequency modulated sine-wave signals with 1\,MHz deviation and 10\,kHz rate; this modulated microwave signal is then amplified and sent to a custom microwave loop antenna to coherently manipulate the NV spin states. 

\begin{figure*}[!htb]
    \centering
    \includegraphics[width=0.8\textwidth]{Fig.8.pdf}
    \caption{\label{sup-fig:ModulationCalibration}
    (a) Demodulated CW-ODMR spectrum from sample S3-14N using frequency-modulated microwave signals. A linear fit is applied to the section indicated by the dash dotted orange line.
    (b) Zoom-in view of the linear fit applied at the zero-crossing point of the left resonance feature in the demodulated CW-ODMR spectrum, yielding slope of $0.591\pm0.007$\,V/MHz.
    (c) FFT spectrum of time-averaged lock-in CW-ODMR measurements, with amplitude modulated $B_{\perp}$ applied with 0.1\,$V_{pp}$ at 100\,Hz. Calibration measurements allow expression of the FFT spectrum in magnetic field units, with peak modulated field amplitude of $10.23 \pm 0.12$\,$\mu$T .
}
\end{figure*}
\noindent

Fig.~\ref{sup-fig:ModulationCalibration}(a) shows an example demodulated CW-ODMR spectrum from measurements on sample S3-14N. A linear slope of $0.591\pm0.007$\,V/MHz of LIA output to microwave frequency is obtained from a linear fit applied to the steepest part of the left NV hyperfine resonance [Fig.~\ref{sup-fig:ModulationCalibration}(b)]. Since the relevant NV spin transition frequencies are sensitive to the projection of magnetic fields along the symmetry axis in the [111] direction, and the bias field used in these lock-in CW-ODMR measurements is in the [$\overline{1}$10] direction, the NV magnetic sensitivity to the modulated $B_\perp$ field is reduced by an angle factor of $\cos(109.47^{\circ}/2) \approx 0.577$. The measured linear slope of LIA output as a function of microwave frequency can then be converted into a calibration constant of $9.56\pm0.12$\,mV/$\mu$T using the NV electronic spin gyromagnetic ratio $\gamma_e = 28.024$\,kHz/$\mu$T and the angle factor. This calibration process is similar to that described in Appendix~\ref{supp:field-calib} for application of bias magnetic fields with the Helmholtz coils.

To avoid the low-pass filter from the LIA affecting the measured signal with $f_m = 100$\,Hz, we set the LIA 3\,dB point at 1\,kHz with a first order filter slope of 6\,dB/octave. For typical LIA measurements, we acquire $n=30$ consecutive 1\,s time traces using a National Instruments USB-6363 data I/O device at a sampling rate of 100\,kHz. The averaged time-domain signal is then transformed into a frequency-domain representation using Fast Fourier Transform (FFT) analysis. The amplitude of the FFT spectrum is converted to the magnetic field unit of $\mu T$ using the calibration constant; and the standard deviation $\sigma$ of the peak amplitudes at 100\,Hz from $n=30$ acquisitions characterizes the measurement uncertainty. Fig.~\ref{sup-fig:ModulationCalibration}(c) depicts an example measured spectrum of the modulated $B_\perp$ field, with peak amplitude of $10.23 \pm 0.12$\,$\mu$T at 100\,Hz determined using the above calibration procedure.

\section{Numerical Simulation Model for AO-PL as a Function of Magnetic Field} \label{supp:Jeyson_model}

To simulate the PL of an ensemble of NV centers with a given NV concentration, we assume that:
(1)  only spin evolution in the electronic ground state needs to be considered, as the time that the system stays in the electronic excited state is much shorter than the time in the ground state; 
(2) optical spin polarization defines the initial condition for evolving the ground spin state; 
(3) evolution of the system is coherent \cite{anishchik_method_2019}; 
(4) the total PL is the average of the steady state PL from all possible pairs of interacting NV centers, with each positioned along four possible crystallographic orientations; 
and (5) the dipolar interaction strength between NVs is fixed by the concentration of the ensemble.

The Hamiltonian that describes the ground state of an isolated NV center $H_{gs}$ is described by:
\begin{equation}
\label{eq:NV-gs-magnetic}
    \frac{\hat{H}_{gs}}{\hbar} = \hat{S}\cdot \textbf{D} \cdot \hat{S} + \hat{S}\cdot \textbf{A} \cdot \hat{I} +  \hat{I}\cdot \textbf{Q} \cdot \hat{I} + \frac{\mu_B}{h}\vec{B}\cdot \textbf{g} \cdot \hat{S}.
\end{equation}
Here, $\textbf{D}$, $\textbf{A}$, $\textbf{Q}$, and $\textbf{g}$ correspond to the fine structure tensor, hyperfine tensor, nuclear electric quadrupole coupling tensor, and g-factor tensor, respectively. 
In addition, $\hat{S}$ and $\hat{I}$ correspond to the NV electronic and nuclear spin operators. $\vec{B}$, $\mu_B$, and $h$ are the bias magnetic field applied uniformly to the system, the Bohr magneton constant, and the Planck constant, respectively.

The magnetic dipolar interaction Hamiltonian $H_{int}$ between two NV centers due to the electronic spin operators $\vec{S}_{1}$ and $\vec{S}_{2}$ is given by: 
\begin{equation}
\label{eq:interaction_term}
    H_{int}= D_{dd}\left[3(\vec{S}_{1}\cdot \hat{n}_{12})(\vec{S}_{2}\cdot \hat{n}_{12})- (\vec{S}_{1} \cdot \vec{S}_{2})  \right].
\end{equation}
$D_{dd}$ describes the dipolar interaction strength, which depends on the relative separation between two NVs, $\vec{r}$. For an NV ensemble with $\sim$3.8\,ppm [NV], $D_{dd} \sim35\,$kHz and scales as $1/|\vec{r}|^3$. 
The unit vector between the two NVs is defined as $\hat{n}_{12}=\vec{r}/|\vec{r}|$. 
For simplicity, we take $D_{dd}$ to be constant for all NV in a given ensemble, as determined by the average NV concentration within the diamond host (assumption (5) above). 

Assuming that the bias magnetic field is aligned along the quantization axis of a single NV orientation class, it is then possible to rewrite Eq.~\ref{eq:NV-gs-magnetic} as: 
\begin{equation}
\label{eq:NV-gs-ham-pas}
\begin{aligned}
    \frac{\hat{H}_{gs}}{\hbar} = & D\left[\hat{S}^2_z - \hat{S}^2/3\right] \\
    + & \frac{\mu_B g_e}{h} \left(B_{NVx}\hat{S}_x + B_{NVy}\hat{S}_y + B_{NVz}\hat{S}_z\right) \\
    + & A^{\parallel}\hat{S}_z\hat{I}_z + A^{\perp}[\hat{S}_x\hat{I}_x +\hat{S}_y\hat{I}_y ] + P\left[\hat{I}^2_z - \hat{I}^2/3\right].
\end{aligned}
\end{equation}
Here, $D$ is the fine structure term called the zero-field splitting (ZFS), $A^{\parallel}$ and $A^{\perp}$ are the axial and transverse hyperfine terms, and P is the nuclear electric quadrupole component. Two important features of the ground states are evident from this Hamiltonian. 
First, the room temperature energy difference between the electronic state ${}^3$A${}_2$ sublevels $\ket{m_s =\pm 1}$ and $\ket{m_s=0}$ is $D(T)\sim 2.87$\,GHz and changes linearly with temperature $dD/dT \approx - 74.2$\,kHz/K \cite{levine_principles_2019-1}. 
Second, the ${}^3$A${}_2$ electronic states have an additional hyperfine energy splitting due to the NV nitrogen nucleus, where $I=1$ and $I=1/2$ correspond to nitrogen isotopes ${}^{14}$N and ${}^{15}$N, respectively. 
The hyperfine parameters are $A^{\parallel}_{14N}\approx -2.14$\,MHz, $A^{\perp}_{14N}\approx -2.70$\,MHz, $P_{14N}\approx -5.01$\,MHz, $A^{\parallel}_{15N}\approx 3.03$\,MHz, $A^{\perp}_{15N}\approx 3.65$\,MHz \cite{anishchik_method_2019,levine_principles_2019-1}. Also, $\gamma_e= g_e \mu_B/h \approx 28.7$ GHz/T is the NV gyromagnetic ratio. The Zeeman interaction lifts the degeneracy between the $\ket{m_s= \pm 1}$ sublevels, i.e., when $|\textbf{B}_{NV}|$ is along an NV quantization axis, the sublevel $\ket{m_s= \pm 1}$ energies split linearly with $|\textbf{B}_{NV}|$, while the $\ket{m_s = 0}$ sublevel energy is unchanged.
\\
\\
In general, Eq.~\ref{eq:interaction_term} can be written as,  

\begin{equation} \label{eq:hamiltonian_decomp}
    H_{Int}=D_{dd}(A+B+C+D+E+F),
\end{equation}

\begin{equation} \label{eq:hamiltonian_A}
A=S_{z}^{1}S_{z}^2\left(1 - 3 \cos{}^2\theta\right),
\end{equation}

\begin{equation}
B=\frac{1}{2}\left(1 - \frac{3}{2} \sin{}^2\theta\right)\left(S_{+}^{1}S_{-}^2+S_{-}^{1}S_{+}^2\right),
\end{equation}

\begin{equation}
C=-\frac{3}{2}\sin \theta \cos \theta e^{-i\phi}\left(S_{z}^{1}S_{+}^2+S_{+}^{1}S_{z}^2\right),
\end{equation}

\begin{equation}
D=-\frac{3}{2}\sin \theta \cos \theta e^{i\phi}\left(S_{z}^{1}S_{-}^2+S_{-}^{1}S_{z}^2\right),
\end{equation}

\begin{equation}
E=-\frac{3}{4}\sin{}^2 \theta  e^{-2i\phi}S_{+}^{1}S_{+}^2,
\end{equation}

\begin{equation} \label{eq:hamiltonian_F}
F=-\frac{3}{4}\sin{}^2 \theta  e^{2i\phi}S_{-}^{1}S_{-}^2.
\end{equation}

Here, $\theta$ and $\phi$ correspond to the polar and azimuthal angles for $\hat{n}_{12}$, respectively. In the above expressions, the term $A$ acts as a local static field, while $B$ leads to flip-flop interactions that produce a change of the electronic spin quantum number of each NV to the opposite value, which conserves the overall system electronic spin quantum number. 
Both of these terms commute with $H_{gs1}+H_{gs2}$, given by Eq.~\ref{eq:NV-gs-ham-pas} for two NVs, since $\Delta m = 0$. 
For this reason, there terms are known as the secular part of the dipole-dipole interaction. 
% They are collectively known as the non-secular part of the Hamiltonian. 
Finally, the full Hamiltonian of two neighboring NV centers in a ensemble is $H= H_{gs1} + H_{gs2}+ H_{int}$. 

We use the von Neumann equation to simulate the PL emission based on the coherent evolution of ground state spins, which can be described as: 
   
\begin{equation}
\label{eq:von_neuman}
    \dot{\rho}=-i \left[H,\rho \right].
\end{equation}
The coupled spin states of the NV electronic spin and nuclear spin for two NVs can be expressed through the following tensor product:
\begin{equation}
  \rho_0 = \rho_{s1}	\otimes  \rho_{s2} \otimes  \rho_{I1} \otimes \rho_{I2}.
\end{equation}

Additionally the initial optical spin polarization can be described by:

\begin{equation}
  \rho_{s1} = \alpha \begin{bmatrix}
0 & 0 & 0\\
0 & 1 & 0 \\
0 & 0 & 0
\end{bmatrix} + \frac{1}{3}\left(1- \alpha\right) \begin{bmatrix}
1 & 0 & 0\\
0 & 1 & 0 \\
0 & 0 & 1
\end{bmatrix},
\end{equation}
where $\alpha$ corresponds to the degree of polarization of the system to $\ket{m_s=0}$. We solve this von Neumman equation in the frame of reference of the full Hamiltonian by diagonalizing the density matrix:

\begin{equation}
    \rho_0^{eb}=\hat{V}^{-1}\rho_0\hat{V}.
\end{equation}

Here, $\hat{V}$ is the unitary matrix formed by the normalized eigenvectors of the full ground state spin Hamiltonian.

We calculate the density matrix in the rotating frame via the following equations:
\begin{equation}
    [H^{eb},\rho^{eb}]= 2\pi \left(E_i - E_j\right)\rho_{ij}^{eb},
\end{equation}

\begin{equation}
\begin{aligned}
    i\frac{d\rho_{ij}^{eb}}{dt} = & 2\pi \left(E_i - E_j\right)\rho_{ij}^{eb} \rightarrow \rho_{ij}^{eb}(t) \\
    = &\rho_{ij}^{eb}(0) \exp \left(-2\pi i \left(E_i - E_j\right)t\right).
\end{aligned}
\end{equation}

We then convert the density matrix back into the lab frame of reference as follows:

\begin{equation}
    \rho(t)=\hat{V}\rho^{eb}(t)\hat{V}^{-1}.
\end{equation}

Next, we calculate the population in the PL bright $\ket{m_s=0}$ state by projecting the above density matrix onto the $\ket{m_s=0}$ spin state as:

\begin{equation}
    \rho_{00}(t)=\text{Tr}\left(P_0 \rho(t)\right).
\end{equation}

We then calculate the average spin population $\rho_{00}$ in the $\ket{m_s=0}$ state using the following expressions:

\begin{equation}
    \langle \rho_{00} \rangle=\text{Tr}\left(P_0 \frac{1}{\tau} \int_{0}^{\infty} \rho(t) \exp\left(-t/\tau\right) dt\right),
\end{equation}

\begin{equation}
\begin{aligned}
    & \frac{1}{\tau} \int_{0}^{\infty} \rho(t) \exp\left(-t/\tau\right) dt \\
    = & \frac{1}{\tau} \int_{0}^{\infty} \hat{V}^{-1} \rho^{eb}(t) \exp\left(-t/\tau\right) \hat{V} dt,
\end{aligned}
\end{equation}

\begin{equation}
    \label{eq:density-matrix-entry}
    \rho_{ij}^{st}= \int_{0}^{\infty}  \rho^{eb}(t) \exp\left(-t/\tau\right) dt = \frac{\rho_{0,ij}^{eb}}{\left(1+2\pi i\left(E_i - E_k)\tau\right)\right)}.
\end{equation}

\begin{equation}
    \label{eq:average-fluorescence}
   \langle \rho_{00} \rangle= \text{Tr} \left( P_0 \hat{V} \rho^{st} \hat{V}^{-1} \right).
\end{equation}

Finally, by using Eqs.~(\ref{eq:density-matrix-entry}, ~\ref{eq:average-fluorescence}) we calculate the AO-PL intensity from two interacting NVs through the full Hamiltonian and a fixed value of the dipolar interaction strength. 
As each NV can be aligned along four possible crystallographic orientations, we describe the total AO-PL intensity from the NV ensemble by averaging the simulated PL from 16 possible configurations of two interacting NVs. 
Numerical simulations of low-field AO-PL intensity use fixed values for hyperfine and other relevant parameters taken as equal to the approximate values given earlier in this section, with the exception of the dipolar interaction strength $D_{dd}$, which is scaled based upon [NV].  The simulated AO-PL contrast is calculated by the normalized difference in simulated intensities \(C = 1 - I_{\text{sig}} / I_{\text{ref}}\) between a simulated AO-PL signal $I_\text{sig}$ and an AO-PL reference $I_\text{ref}$ determined from the maximal simulated AO-PL intensity for a given scan of $B_{\parallel}$. This normalization process is similar to the calculation of experimental AO-PL contrast as described in Sec.~\ref{sec:full-spectrum} of the main text.

\section{Hamiltonian Analysis of AO-PL Features}
\label{supp:Jeyson_analysis}
Now we discuss the physical origin of the various AO-PL contrast features at different magnetic fields given by the numerical simulation using the model presented in the previous section. 

Taking advantage of the additivity of the von Neumman equation, the model allows us to study the effect of each component of the Hamiltonian by isolating the associated terms.
We decompose the simulated AO-PL spectrum to understand the contributions from pairs of NV centers along various crystallographic orientations.

\begin{figure*}[!htb]
    \centering
    \includegraphics[scale=0.9]{Fig.9.pdf}
  \caption{\label{sup-fig:decomposition} Numerical simulation of low-field AO-PL contrast as a function of applied magnetic fields, from: (a) dipole-dipole electronic interactions only (i.e., no hyperfine); (b) full Hamiltonian including hyperfine interactions, with clusters of linear AO-PL contrast features labeled for the first quadrant; (c) only secular terms $H_A + H_B$; (d) only $H_C + H_D$; and (e) only $H_E + H_F$.}
\end{figure*}
\noindent

\begin{figure*}[!htb]
    % \label{fig:supp_decomposition}
    \centerline{\includegraphics[scale=0.8]{Fig.10.pdf}}

\caption{\label{fig:decomposition-planes} Decomposition of simulated AO-PL contrast as a function of applied magnetic fields, for specifically oriented interacting NV center pairs. (a) and (b) show the contributions from the secular terms ($H_A + H_B$) and non-secular terms ($H_E + H_F$), respectively. Each panel contains 16 sub-figures, corresponding to decompositions for NV pairs aligned along one of the four possible crystallographic orientations, as defined in the main text.}

\end{figure*}

Fig.~\ref{sup-fig:decomposition}(b) shows the numerically simulated low-field AO-PL contrast spectrum for the full Hamiltonian, averaged over all 16 pairs of NV orientations and for values of $B_{\parallel}$ and $B_{\perp}$ similar to that used in this study's experiments. The simulation shows different clusters of linear AO-PL contrast features; those in the first quadrant are labeled as (I), (II), (III), (IV) and (V).
We start with simulations that remove the hyperfine interaction (i.e., NV electronic interactions only), as shown in Fig.~\ref{sup-fig:decomposition}(a). 
The number of lines (linear AO-PL contrast features) within each cluster are reduced in this case, while the normalized AO-PL contrast is higher. This result is consistent with the NV hyperfine interaction introducing additional cross-relaxation channels for the NV spin population, including various shoulder AO-PL features as discussed in the main text. 

Figs.~\ref{sup-fig:decomposition}(c), (d), and (e) show simulated AO-PL contrast for only $H_A+H_B$, $H_C+H_D$, and $H_E+H_F$ in Eq.~\ref{eq:hamiltonian_decomp}, respectively. 
When the secular terms $H_A+H_B$ are considered [Fig.~\ref{sup-fig:decomposition}(c)], there are no AO-PL contrast features within cluster (III), and the number of lines in clusters (II) and (V) are reduced. 
Terms $H_C+H_D$ describe magnetic-field-dependent spin-mixing and associated AO-PL contrast features [Fig.~\ref{sup-fig:decomposition}(d)], with no NV-NV cross-relaxation signals.
Terms $H_E+H_F$ describe double spin-flip processes at specific magnetic field angles [Fig.~\ref{sup-fig:decomposition}(e)] that contribute to the remaining AO-PL features. 

We next consider the AO-PL contrast features from the perspective of constraints placed by the interaction Hamiltonian (Eq.~\ref{eq:hamiltonian_decomp}) on $m$ and $m'$, the magnetic moment quantum numbers for the electronic spins  of the two interacting NVs  in our model system. These quantum numbers are affected by each term of the interaction Hamiltonian (Eqs.~\ref{eq:hamiltonian_A} to \ref{eq:hamiltonian_F}) in the following way:

\[
\begin{aligned}
A &: \Delta m = 0, \quad \Delta m' = 0, \quad \Delta(m + m') = 0, \\
B &: \Delta m = \pm 1, \quad \Delta m' = \mp 1, \quad \Delta(m + m') = 0, \\
C &: \Delta m = 
\begin{cases} 
0, \\ 
1, 
\end{cases}
\quad \Delta m' = 
\begin{cases} 
1, \\ 
0, 
\end{cases}
\quad \Delta(m + m') = 1, \\
D &: \Delta m = 
\begin{cases} 
0, \\ 
-1, 
\end{cases}
\space \Delta m' = 
\begin{cases} 
-1, \\ 
0, 
\end{cases}
\space \Delta(m + m') = -1, \\
E &: \Delta m = 1, \quad \Delta m' = 1, \quad \Delta(m + m') = 2, \\
F &: \Delta m = -1, \quad \Delta m' = -1, \quad \Delta(m + m') = -2.
\end{aligned}
\]

Term \(A\) has the same form as two classical interacting dipoles and describes the effect of a static local field. 
Term \(B\) allows for a simultaneous flip of two neighboring NV electronic spins in opposite directions while keeping the total electronic spin quantum number of the system (S1 + S2) unchanged; this `flip-flop' term corresponds to the resonance effect of the local field from the NV-NV dipolar interaction.

The contributions of terms \(C\) and \(D\) make the energies of the two NVs not degenerate, and hence do not introduce any NV-NV spin cross-relaxation and associated AO-PL contrast. 

Finally, terms \(E\) and \(F\) describe a simultaneous electronic spin flip from both NVs. This `double spin-flip' process is the origin of cluster (III) in Fig.~\ref{sup-fig:decomposition}(b) and contributes to additional shoulder features in clusters in (V) and (II). 

Further decomposition of the contributions from specific interaction terms can be performed by considering pairs of interacting NV classes separately, providing a clearer understanding of the nature of each AO-PL contrast feature, as shown in Fig.~\ref{fig:decomposition-planes}. 
For instance, this analysis indicates that vertical cross-relaxation lines (I) in Fig.~\ref{sup-fig:decomposition}(b) arise from flip-flop interactions driven by off-axis ($\hat{n}_{\phi}, \hat{n}_{\kappa}, \hat{n}_{\chi}$) NV pairs belonging to different NV classes. 
Horizontal lines (V) in Fig.~\ref{sup-fig:decomposition}(b) result from contributions of both flip-flop and double spin-flip interactions, involving same- and different-class NV pairs from $\hat{n}_{\lambda}$ and $\hat{n}_{\phi}$. Additionally, a double spin-flip interaction occurs between an NV pair where one center belongs to the $\hat{n}_{\chi}$ class and the other to the $\hat{n}_{\kappa}$ class. This suggests that certain peaks along these cross-relaxation lines are specifically linked to flip-flop, double spin-flip, or a combination of both interactions.

On the other hand, examination of Fig.~\ref{fig:decomposition-planes} indicates that cross-relaxation lines (II) in Fig.~\ref{sup-fig:decomposition}(b) originate from flip-flop and double spin-flip contributions within the same NV class, specifically from $\hat{n}_{\kappa}$ and $\hat{n}_{\chi}$. 
Lines (IV) in Fig.~\ref{sup-fig:decomposition}(b) appear when on-axis NVs $\hat{n}_{\lambda}$ interact with off-axis NVs in the $n_{\chi}$ and $n_{\kappa}$ directions via flip-flop interactions. 
Finally, lines (III) in Fig.~\ref{sup-fig:decomposition}(b) are due to the double spin-flip terms, arising from pairs of interacting NV centers oriented along $\hat{n}_{\lambda}$ and $\hat{n}_{\chi}$; $\hat{n}_{\phi}$ and $\hat{n}_{\kappa}$.

The insight that some AO-PL contrast features are predominantly governed by either flip-flop or double spin-flip transitions opens opportunities to further explore the underlying physics, e.g., the effect of applied electric fields, which have been shown to enhance double spin-flip transition rates in previous studies \cite{pellet-mary_cross-relaxation_2022}. 
Additionally, the ability to isolate different NV-class contributions in the AO-PL feature map can be further leveraged in combination with techniques such as Fourier imaging \cite{arai_fourier_2015, backlund_fourier_2017}, enabling targeted analysis of specific NV classes.

\begin{figure*}[!htb]
    \centering
    \includegraphics[width=0.7\textwidth]{Fig.11.pdf}
    \caption{\label{sup-fig:concentrationComparison}
    (a) (Top) Experimentally measured AO-PL contrast as a function of applied magnetic fields using sample S5-14N with $\approx \,$0.3\,ppm [NV]. Weak NV-NV cross-relaxation features are observed in addition to the background contrast from magnetic-field-dependent spin mixing. (Bottom) Simulated AO-PL contrast using a fixed dipolar interaction strength from 0.3\,ppm [NV]. Both spin mixing and NV-NV cross-relaxation AO-PL contrast features are seen.
    (b) (Top) Experimentally measured AO-PL contrast as a function of applied magnetic fields using sample S1-14N with $\approx \,$3.8\,ppm [NV]. (Bottom) Simulated AO-PL contrast using a fixed dipolar interaction strength from 3.8\,ppm [NV].
}
\end{figure*}

\section{AO-PL Spectra with Different NV Concentrations}
\label{supp:spectra-concentration}

Figs.~\ref{sup-fig:concentrationComparison}(a) and \ref{sup-fig:concentrationComparison}(b) show the measured (top) and simulated (bottom) AO-PL contrast as a function of $B_\parallel$ and $B_\perp$ from sample S5-14N with $\approx \,$0.3\,ppm NV concentration and sample S1-14N with $\approx \,$3.8\,ppm NV concentration, respectively. Clusters of linear AO-PL contrast features arising from NV-NV cross-relaxation are experimentally observed near zero-field, $\theta = 0$, and $\theta = 39.3$\textdegree\ for both the low and high NV concentration samples, which have relatively weak and strong NV-NV dipolar interaction strength, respectively. In addition, there is a non-negligible AO-PL contrast background, increasing with bias field magnitude, due to the mixing of NV electronic spin states from magnetic fields that are not aligned with the NV quantization axis \cite{tetienne_magnetic-field-dependent_2012}. Future work may incorporate field-dependent transition rates between spin sublevels and thereby improve simulations of AO-PL contrast from both spin mixing and NV-NV cross-relaxation.

\section{AO-PL as a Function of Laser Power}
\label{supp:AOPL_power}
\subsection{AO-PL Linewidth}
\label{supp:linewidth_laser}
As described in Sec.~\ref{sec:nuclear-hyperfine}, the ratio of individual AO-PL contrast peak amplitudes using a $^{14}$N-enriched sample can be described as 1:2:3:2:1, at $B_\parallel=1.24\,$mT and $B_\perp \approx 1.05\,$mT [Fig.~\ref{fig:14vs15}(b, top)]. We model the observed quintet of peaks as a function of $B_\perp$ using the following multi-Lorentzian function $L(x)$:
    \begin{align} 
        L(x) = & \frac{A_1}{\left[1 + \frac{(B_\perp - B_1)}{(\gamma/2)}^2\right]} 
        + \frac{2A_1}{\left[1 + \frac{(B_\perp - (B_1 + \Delta B))}{(\gamma/2)}^2\right]} \nonumber \\
        + & \frac{3A_1}{\left[1 + \frac{(B_\perp - (B_1 + 2\Delta B))}{(\gamma/2)}^2\right]}
        + \frac{2A_1}{\left[1 + \frac{(B_\perp - (B_1 + 3\Delta B))}{(\gamma/2)}^2\right]} \nonumber \\
        + & \frac{A_1}{\left[1 + \frac{(B_\perp - (B_1 + 4\Delta B))}{(\gamma/2)}^2\right]}
        + kB_\perp \label{eqn:supp-multiLorentzian}
    \end{align}
Here, $A_1$ and $B_1$ describe the left-most peak's AO-PL contrast and resonance location, respectively. We restrict the fit model by assuming all peaks have the same linewidth $\gamma$ and separation $\Delta B$. A change of baseline is seen in the experimental data between $B_\perp \approx$ 0.8 and 1.4\,mT due to magnetic-field-dependent spin mixing, which we model as a linear shift $kB_\perp$. The fits are done in MATLAB using the nonlinear curve-fitting toolbox with a Levenberg-Marquardt algorithm. 

Fig.~\ref{sup-fig:linewidth_laserPower_FIT} shows the measured AO-PL contrast and the corresponding model fit as a function of $B_\perp$, at fixed $B_\parallel = 1.24\,$mT, for sample S3-14N at (a) 2\,mW and (b) 50\,mW laser excitation power. At the higher laser power, we observe a decrease of AO-PL contrast; also the shift of contrast baseline from spin-mixing becomes more prominent.
Note that the fit to experimental data is used to determine the linewidth, peak separation, and linear baseline shift (i.e., $\gamma$, $\Delta B$, and $k$ are free fit parameters) from this quintet AO-PL feature. 
AO-PL contrast at the central peak is calculated from the ratio of a measurement at $B_\perp = 1.05$\,mT and a reference measurement at $B_\parallel = 0.73$\,mT, both at $B_\parallel = 1.24$\,mT as described in Sec.~\ref{sec:contrast_laser} in the main text.

\begin{figure*}[!htb]
    \centering
    \includegraphics[width=0.75\textwidth]{Fig.12.pdf}
    \caption{\label{sup-fig:linewidth_laserPower_FIT}
    Experimentally measured AO-PL contrast as a function of $B_\perp$ and at fixed $B_\parallel = 1.24\,$mT (dotted markers) with multi-Lorentzian fits (solid lines) for sample S3-14N at (a) 2\,mW and (b) 50\,mW laser excitation power. Fit values for the AO-PL contrast peak linewidth $\gamma$ are shown. A linear shift of baseline due to magnetic-field-dependent spin mixing becomes more prominent at higher laser powers accompanied by a reduction of AO-PL contrast. 
}
\end{figure*}
\noindent

\begin{figure}[!htb]
    \centering
    \includegraphics[width=0.48\textwidth]{Fig.13.pdf}
    \caption{\label{sup-fig:linewidth_laserPower}
    Extracted AO-PL contrast feature linewidths $\gamma$ from fits to measurements around $B_\perp \approx\,$1\,mT using samples S3-14N (blue) and S4-14N (orange) as a function of laser power and intensity. The laser intensity is calculated by the same procedure described in Fig.~\ref{fig:laser_dependency}. Each sample exhibits broader linewidth at higher laser power. Sample S3-14N has higher [NV] and significantly larger linewidth than sample S4-14N.
}
\end{figure}
\noindent

Fig.~\ref{sup-fig:linewidth_laserPower} shows the AO-PL contrast peak feature linewidth $\gamma$, determined from fits of Eq.~\ref{eqn:supp-multiLorentzian} to measurements such as in Fig.~\ref{sup-fig:linewidth_laserPower_FIT}, as a function of laser power using samples S3-14N and S4-14N with varying [NV] ($\approx\,$3.8\,ppm and 2\,ppm, respectively). Sample S3-14N with higher [NV], and hence stronger NV-NV dipolar interaction, exhibits a roughly 50\% greater linewidth than sample S4-14N across all laser powers studied. 

Our results of linewidth as a function of laser power deviate from the reported linewidth narrowing with increased laser power using nanodiamonds in Ref.~\cite{dhungel_near_2024}.

\subsection{Modeling AO-PL Contrast Changes from Optical Pumping and Resonant NV-NV Interaction}
\label{supp:contrast_competition}
Appendix~\ref{supp:Jeyson_model} describes the use of a density matrix approach to calculating the AO-PL contrast from an NV ensemble, with two NV centers at fixed dipolar interaction strength that are then averaged over all possible pairs of crystallographic orientations. 
This method provides reasonable agreement with low-field AO-PL experimental results [Figs.~\ref{fig:concentration_spectrum}(b, c), \ref{sup-fig:concentrationComparison}].
However, the model only considers the initial spin polarization without accounting for details of the optical pumping process.
To understand the dependence on laser power of the maximum AO-PL contrast from the quintet peaks, as discussed in Sec.~\ref{sec:contrast_laser} in the main text, we now describe an alternative numerical model using rate equations for a single NV in one of the four orientations in the diamond host. We find results from this rate-equation model are in reasonable agreement with measurements, including AO-PL contrast line features as a function of applied magnetic field and AO-PL contrast as a function of laser power.

For the diamond samples in the present study, the average nearest-neighbor NV-NV dipole interaction $\sim$30\,kHz, which is much smaller than the energy separation between NVs of different orientations when their electronic spin transitions (with typical linewidth $\sim$1\,MHz) are spectrally separated. Therefore, we first proceed by diagonalizing the ground state Hamiltonian described in Eq.~\ref{eq:NV-gs-magnetic} without adding the dipole-dipole interaction terms (Eq.~\ref{eq:interaction_term}), and then calculate the steady-state AO-PL by solving the resulting rate equations. 
The total AO-PL is determined by summing over all possible NV orientations, similar to the procedure described in Appendix~\ref{supp:Jeyson_model}.
In the presence of near-degenerate transition frequencies, e.g., induced for particular on ($B_\parallel$) and off ($B_\perp$) axis bias magnetic fields or by hyperfine interactions, the decay rates between various spin states increase.

We simplify a 5-level spin population model \cite{robledo_spin_2011, barry_sensitivity_2020} of a single NV –– which includes ground and excited electronic states, as well as a single metastable state representing the singlet state manifold, as shown in Fig.~\ref{fig:Schematics}(a) –– to a 4-level system by integrating out the intersystem crossing singlet state, as diagrammed in Fig.~\ref{sup-fig:Mike_model}(a). 
Given that we perform steady-state operations in the experiment, and the singlet state is a short-lived intermediate state with a lifetime $\approx$ 300\,ns, we model the transitions through this state via an effective radiative decay from electronic excited states to the ground states. 

\begin{figure*}[!htb]
    \centering
    \includegraphics[width=0.7\textwidth]{Fig.14.pdf}
    \caption{\label{sup-fig:Mike_model}
    (a) Simplified 4-level system used in the rate-equation model by integrating out the intersystem crossing singlet state manifold shown in Fig.~\ref{fig:Schematics}(a) in the main text. Relaxation rates are included between ground ($\Gamma$) and excited ($D$) spin sublevels.
    (b) Simulated AO-PL contrast as a function of $B_\perp$ at fixed $B_\parallel = 1.24\,$mT using the 4-level population model provides comparable results to the density matrix method described in Appendix~\ref{supp:Jeyson_model}.
}
\end{figure*}
\noindent

In addition, we introduce relaxation rates between the $\ket{\pm 1}$ and $\ket{0}$ spin sublevels in both the electronic ground ($\Gamma$) and excited states ($D$). Therefore, the model 4-level system can be described with the following rate equations: 

\begin{equation}
\begin{aligned}
  & \partial_t e_0 = B(n_0-e_0)-\gamma e_0 + D(\beta' e_1-(1-\beta')e_0), \\
  & \partial_t e_1 = B(n_1-e_1)-\gamma' e_1 - D(\beta' e_1-(1-\beta')e_0), \\
  & \partial_t n_0 = -B(n_0-e_0) +\gamma_a e_0 +\gamma_b e_1 \\
  & \qquad \quad+ \Gamma(\beta n_1-(1-\beta)n_0), \\
  & \partial_t n_1 = -B(n_1-e_1)+\gamma'_a e_1 +\gamma'_b e_0 \\
  & \qquad \quad- \Gamma(\beta n_1-(1-\beta)n_0).
\end{aligned}
\end{equation}

Here, $n_{0,1}$ ($e_{0,1}$) refer to the population in the electronic $\ket{0}$ and $\ket{\pm 1}$ ground states (excited states, respectively). 
The parameter $B$ is proportional to the optical pumping intensity.
$\beta' = \beta=\frac{1}{3}$ sets the equilibrium population distribution across the $\ket{0}$ and $\ket{\pm1}$ states without optical excitation. $\gamma+\gamma' = \gamma_a+\gamma_b+\gamma'_a+\gamma'_b$ are the net excited state decay rates into the ground states that arise from integrating out the singlet state, where $\gamma$ ($\gamma'$) is the effective decay rate from the excited $\ket{0}$ (respectively $\ket{\pm1}$) states; $\gamma_a$ and $\gamma_b$ ($\gamma_a'$ and $\gamma_b'$) are the decay rates towards the ground $\ket{0}$ (respectively ground $\ket{\pm1}$) states. 
We follow the summary of experimental NV results given in Ref.~\cite{barry_sensitivity_2020} and use the approximate expressions $\gamma = s+s'/7$,  $\gamma'=s+s'$, where $s$ is the direct radiative decay rate from the excited state manifold and $s'$ is the decay rate from the excited state to the singlet state, with $s = s'$ = $6.7 \times 10^7 s^{-1}$. 
Integrating out the singlet intermediate state yields $\gamma_a = s+2s'/21$, $\gamma_b = 2s'/3$, $\gamma'_a = s+s'/3$ and $\gamma'_b = s'/21$. 
The AO-PL intensity is finally calculated as being proportional to the excited state populations in each electronic spin state. 
An example of simulated AO-PL contrast as a function of $B_\perp$ at fixed $B_\parallel = 1.24\,$mT is shown in Fig.~\ref{sup-fig:Mike_model}(b).

Ref.~\cite{choi_depolarization_2017} indicates that the NV spin depolarization rate increases when sub-ensemble spin transitions become degenerate, and samples with larger [NV] have shorter NV ground state spin polarization lifetime ($T_1$).
Thus, we associate the ground state spin relaxation rate $\Gamma$ and the excited state spin relaxation rate $D$ introduced in the 4-level model [Fig.~\ref{sup-fig:Mike_model}(a)] to depend on both spin transition degeneracy and NV concentration. 
Spin transition degeneracy can lead to maximal state mixing between different NV classes, with enhanced NV-NV cross-relaxation and hence reduction in AO-PL. 
We model $\Gamma$ to include a term that is proportional to [NV]$^2$, as expected from Fermi's golden rule and associated squaring of the matrix element for the dipolar interaction strength (see Eq.~\ref{eqn:int} in main text and also \cite{choi_depolarization_2017}). The resulting ground state spin relaxation rate $\Gamma$ is parameterized as:

\begin{equation}
  \Gamma = \Gamma_0 + \kappa [NV]^2 (1+d).
\end{equation}
Here $\Gamma_0$ is the single NV ground state spin relaxation rate, $\kappa$ is a parameter calibrated by comparing simulation with experiment, and $d$ is a scaled dipolar interaction strength characterizing NV-NV cross-relaxation. We use an analogous expression for $D$, the excited state spin relaxation rate.

Appendix~\ref{supp:T1_measurement} describes the procedures to experimentally extract the ground state $T_1$ lifetime at different magnetic field configurations. The coefficient $\Gamma|_{d=0}$ is given by the inverse of this measured $T_1$ lifetime when there are no spin state degeneracies.
$\kappa$ and $d$ are indirectly determined from the measured ground state $T_1$ lifetime as a function of magnetic field configuration ($B_\parallel$ and $B_\perp$). 
For example, for sample S3-14N ($\approx$ 3.8\,ppm [NV]) we measure the ground state spin relaxation rate ($\Gamma$) $\approx$ 0.21 $\pm$ 0.005\,$\text{ms}^{-1}$ at magnetic field configurations where NV-NV cross-relaxation occurs (i.e., at linear AO-PL contrast features). 
When all NV ground state spin resonances are spectrally well-separated, we measure the spin relaxation rate ($\Gamma|_{d=0}$) $\approx$ 0.18 $\pm$ 0.005\,$\text{ms}^{-1}$. 
By incorporating a Gaussian excitation laser beam profile of radius 33\,$\mu m$, we then determine the associated parameters for $D$, the excited state spin relaxation rate, by numerical fit of the 4-level model to the measured AO-PL contrast as a function of laser power for this sample, as shown by the blue solid line in Fig.~\ref{fig:laser_dependency} in the main text. 

Taking these results and scaling [NV] down by a factor of 1.35$\times$, we find the simulated AO-PL contrast as a function of laser power to be in reasonable agreement with the measured results from sample S4-14N ($\approx$ 2\,ppm [NV], orange solid line in Fig.~\ref{fig:laser_dependency} in the main text). 
This scaling factor is roughly consistent with the $\sim 2 \times$ [NV] ratio between sample S3-14N and S4-14N inferred from the variations of irradiation dose during electron implantation \cite{edmonds_characterisation_2021}.

Table~\ref{supp-tab:model_values} in the Appendix summarizes the measured and inferred relaxation rates between ground state sublevels $\Gamma$ and excited state sublevels $D$ used to generate the model-based curves for samples S3-14N and S4-14N shown in Fig.~\ref{fig:laser_dependency} in the main text. 
$\Gamma$ and $D$ apply to magnetic field configurations where AO-PL contrast features are maximally degenerate;
$\Gamma|_{d=0}$ and $D|_{d=0}$ refers to configurations where all electronic spin states are spectrally well-separated. 

\begin{table*}
\centering
\caption{\label{supp-tab:model_values} Relaxation rates between NV spin sublevels, both with ($\Gamma$ and $D$) and without ($\Gamma|_{d=0}$ and $D|_{d=0}$) spin-state degeneracies, as used in the 4-level rate equation simulation model. Values with * are determined experimentally. Other values are inferred from measurements as described in Appendix~\ref{supp:contrast_competition}.}
\begin{ruledtabular}
% \resizebox{\textwidth}{!}{%
\begin{tabular}{ccccc}
\textbf{Sample \#}  & \textbf{$\Gamma$ (ms$^{-1}$)} & \textbf{$\Gamma|_{d=0}$ (ms$^{-1}$)} & \textbf{{$D$ (ms$^{-1}$)}} & \textbf{$D|_{d=0}$ (ms$^{-1}$)}  \cr
S3-14N                       & $0.21\pm0.005$*    &  $0.18\pm0.005$*             &249  & 7.1 \cr
S4-14N                       & 0.13    &  0.12           &119  &  2.1 \cr
\end{tabular}
\end{ruledtabular}
\end{table*}

We associate the observed initial increase of AO-PL contrast at laser power $<1\,$mW (Fig.~\ref{fig:laser_dependency} in the main text) to improved NV electronic spin polarization from optical pumping; and  attribute the subsequent decrease of contrast at higher laser powers to competition between increased net optical pumping and depolarization from NV-NV cross-relaxation. 
By taking $\Gamma$ and $D$ to be both [NV] and spectral degeneracy dependent, our model reproduces the experimentally observed behavior that the maximum AO-PL contrast for sample S3-14N with larger [NV] occurs at a higher laser power ($\approx 0.95\,$mW) than for sample S4-14N ($\approx 0.55\,$mW). 
Further study will be necessary to fully delineate the physical processes most relevant for the observed AO-PL contrast behavior.

\begin{figure*}[!htb]
    \centering
    \includegraphics[width=0.7\textwidth]{Fig.15.pdf}
    \caption{\label{sup-fig:T1}
    Measurement of sample S3-14N polarization lifetime ($T_1$) under different magnetic field configurations.
    (a) Experimental sequence used to measure $T_1$.
    (b) CW-ODMR measurement with spectrally well-separated spin resonances for all NV orientations ($B_\parallel \approx 2.4\,$mT and $B_\perp \approx 1.1\,$mT).
    (c) Measured peak amplitude of leftmost ODMR resonance in (b) as a function of delay time $\tau$ shown in experimental sequence in (a). Fit of decay yields $T_1= 5.46\pm0.05\,$ms. $T_1$ measurements and fits are repeated for all ODMR resonances in (b), giving averaged $\overline{T_1}$ = $5.51\pm0.15$\,ms.
    (d) CW-ODMR measurement with degenerate spin resonance for NV orientations $\hat{n}_\lambda$ and $\hat{n}_\chi$; $\hat{n}_\phi$ and $\hat{n}_\kappa$.
    (e) Measured peak amplitude of leftmost ODMR resonance in (d) as a function of $\tau$; fit yields $T_1= 4.72\pm0.04\,$ms. $T_1$ measurements and fits are repeated for all ODMR resonances (d), giving averaged $\overline{T_1}$ = $4.74\pm0.1$\,ms.
}
\end{figure*}
\noindent

\subsection{Polarization Lifetime ($T_1$) Measurements on Sample S3-14N}
\label{supp:T1_measurement}
Fig.~\ref{sup-fig:T1}(a) shows the experimental sequence employed to measure the NV ground state spin polarization lifetime ($T_1$) of sample S3-14N as a basis for the 4-level population model described in Appendix~\ref{supp:contrast_competition}.
NV electronic spins are first optically polarized by an initial laser pulse to (mostly) ground spin state $\ket{0}$; and are then optically read out after a variable dark time $\tau$.
In the second half of the sequence, as a reference measurement, a microwave $\pi$ pulse (orange) prepares the NV electronic spin to $\ket{+1}$ or $\ket{-1}$.
By dividing the two results, the NV spin polarization $P(t)$ can be expressed as \(P(t) = e^{-t/T_1}\).

Fig.~\ref{sup-fig:T1}(b) presents an example CW-ODMR measurement where all NV orientation classes are spectrally well-separated (with $B_\parallel \approx 2.4\,$mT and $B_\perp \approx 1.1\,$mT). Employing the measurement sequence in Fig.~\ref{sup-fig:T1}(a), a fit of P(t) to the peak amplitude of the leftmost ODMR resonance feature as a function of $\tau$ yields $T_1 = 5.46\pm0.05\,$ms [Fig.~\ref{sup-fig:T1}(c)]. 
We repeat these measurements and fitting procedure for the peak amplitudes of all ODMR resonances in Fig.~\ref{sup-fig:T1}(b); and then average the fit $T_1$ values for all resonances and obtain $\overline{T_1} = 5.51\pm0.15\,$ms, equivalent to an NV spin relaxation rate with no spin-state degeneracies $\Gamma|_{d=0} = 0.18\pm0.005\,\text{ms}^{-1}$.

Next, we change the bias magnetic field so that the pair of NVs along $\hat{n}_\lambda$ and $\hat{n}_\chi$ and the pair of NVs along $\hat{n}_\phi$ and $\hat{n}_\kappa$ have degenerate spin resonances, with $B_\parallel = 1.24\,$mT and $B_\perp \approx 1.05\,$mT.
An example CW-ODMR measurement for this bias field configuration is shown in Fig.~\ref{sup-fig:T1}(d). Again employing the measurement sequence in Fig.~\ref{sup-fig:T1}(a), we find $\overline{T_1}$ = $4.74\pm0.1$\,ms [Fig.~\ref{sup-fig:T1}(e)], equivalent to an NV spin relaxation rate with spin-state degeneracies $\Gamma = 0.21\pm0.005\,\text{ms}^{-1}$.

\section{Contribution of Hyperfine Splitting to Separation of AO-PL Contrast Peaks}
\label{supp:hyperfine_separation}
To determine the splitting $\Delta B_{15N}$ from the measured triple AO-PL contrast peaks using sample S2-15N (see Fig.~\ref{fig:14vs15} and \ref{fig:15N_lockin} of the main text), we perform a multi-Lorentzian fit to the data that is similar to $L(x)$ in Eq.~\ref{eqn:supp-multiLorentzian}, except that there are only three peak amplitudes following a 1:2:1 ratio. As shown in Fig.~\ref{sup-fig:15N_peakSeparation}(a), the fit yields $\Delta B_{15N, \text{AO–PL}} = 0.1289 \pm 0.0002$\,mT. 

\begin{figure*}[!htb]
    \centering
    \includegraphics[width=0.85\textwidth]{Fig.16.pdf}
    \caption{\label{sup-fig:15N_peakSeparation}
    (a) Measured AO-PL contrast as a function of $B_\perp$ (dotted markers) with multi-Lorentzian fit (solid line) for sample S2-15N and fixed $B_\parallel = $1.24\,mT. The extracted separation between AO-PL contrast peaks $\Delta B_{15N, \text{AO–PL}}$ = 0.1289 $\pm$ 0.0002\,mT. (b) Linear fit applied to CW-ODMR transition frequencies as a function of $B_\perp$ for NV orientations $\hat{n}_\chi$ and $\hat{n}_\kappa$, from measurements shown in Fig.~\ref{fig:15N_lockin}(c, right) of the main text with fixed $B_\parallel = $1.24\,mT. The linear slope $m = 23.568 \pm 0.034$\,MHz/mT can be treated as the effective gyromagnetic ratio of the NV electronic spin for $B_\perp$ along [$\overline{1}$10].
}
\end{figure*}
\noindent

Since $B_\perp$ is transverse to both $\hat{n}_\lambda$ and $\hat{n}_\phi$ for these measurements, the spin transition frequencies from the NVs along either orientation should remain the same as $B_\perp$ is varied, separated by the associated nitrogen nuclear spin hyperfine splitting. In contrast, $B_\perp$ has non-zero projection along NV orientations $\hat{n}_\chi$ and $\hat{n}_\kappa$. Thus, the magnetic field separation between neighboring AO-PL peaks and CW-ODMR spin transition frequencies for these orientations can be calculated through consideration of the hyperfine splitting $A_N$, the NV electronic spin gyromagnetic ratio $\gamma_e$, and the angle factor $\cos(\alpha)$ from the projection of $B_\perp$ along each NV orientation, as described in main text Sec.~\ref{sec:nuclear-hyperfine}, yielding \(\Delta B_N = A_N / [\gamma_e \cos(\alpha)]\). The linear slope of the measured CW-ODMR transition frequencies from NVs along $\hat{n}_\chi$ and $\hat{n}_\kappa$, shown in Fig.~\ref{fig:15N_lockin}(c, right), can then be associated to $\gamma_e \cos(\alpha)$. Fig.~\ref{sup-fig:15N_peakSeparation}(b) shows the fit slope from these CW-ODMR transition frequencies as a function of $B_\perp$, giving $m = 23.568 \pm 0.034$\,MHz/mT. For $^{15}$N, the hyperfine splitting $A_{15N}$ $\approx\,$3.03\,MHz. Therefore, we calculate the CW-ODMR triple feature splitting $\Delta B_{15N, \text{CW–ODMR}} = 0.1286 \pm 0.0002$\,mT. The good agreement between $\Delta B_{15N, \text{AO–PL}}$ and $\Delta B_{15N, \text{CW–ODMR}}$ confirms that NV hyperfine interactions contribute to the observed AO-PL signals, including the separation between AO-PL contrast peaks.
In addition, we can extract the AO-PL contrast feature linewidth $\gamma = 58.34 \pm 0.66 \,\mu T$ through the same fit shown in Fig.~\ref{sup-fig:15N_peakSeparation}(a). By using the fit slope value $m$ from Fig.~\ref{sup-fig:15N_peakSeparation}(b), this linewidth in magnetic field units can be converted to $\gamma' = 1.375 \pm 0.016$\,MHz in frequency units.

\section{AO-PL Line Shape Comparisons Between Near-Zero-field and Low-Field}
\label{supp:slope_comparison}
Fig.~\ref{sup-fig:zeroLowField} depicts the measured AO-PL contrast as a function of $B_\perp$ using samples S3-14N and S4-14N, which differ by [NV] (see Table~\ref{tab:samples} in the main text). At $B_\parallel = 1.24$\,mT and near $B_\perp = 1.05$\,mT, Fig.~\ref{sup-fig:zeroLowField}(b, d), the AO-PL contrast feature linewidth $\gamma$ is extracted using the same Lorentzian fit function as described in Appendix~\ref{supp:linewidth_laser} by constraining the peak contrast to follow 1:2:3:2:1. 
For the fit of data near zero-field ($B_\parallel = 0.05\,$mT), Fig.~\ref{sup-fig:zeroLowField}(a, c), due to the increased number of spectral degeneracies, we set the amplitudes of the left three AO-PL contrast peaks as free parameters, while the amplitudes of the remaining two peaks on the right are constrained to mirror those on the left. For these fits, we find that the linewidth $\gamma$ decreases by $\sim2.3\times$ (from 119.3\,$\mu T$ to 50.9\,$\mu T$) and $\sim2.5\times$ (from 83.5\,$\mu T$ to 33.5\,$\mu T$) when comparing near-zero-field to low-field, using samples S3-14N and S4-14N, respectively. The increased linewidth near zero-field could be due to NV spin-state mixing induced by local electric fields \cite{pellet-mary_relaxation_2023} and/or inhomogeneities in the system. We note that the misalignment of magnetic fields near zero-field can easily broaden the linewidth by enhancing overlap of NV spin spectral features and making hyperfine structures harder to resolve. 

\begin{figure*}[!htb]
    \centering
    \includegraphics[width=0.9\textwidth]{Fig.17.pdf}
    \caption{\label{sup-fig:zeroLowField}
    Measured AO-PL contrast as a function of $B_\perp$ (dotted markers) with multi-Lorentzian fits (solid lines) using samples (a, b) S4-14N and (c, d) S3-14N. Left: results near zero-field ($B_\parallel = 0.05\,$mT). Right: results near $B_\parallel = 1.24$\,mT and $B_\perp = 1.05$\,mT. Both samples exhibit AO-PL contrast feature linewidths $\sim2.5\times$ broader than values at low-field. Sample S4-14N has $\sim2\times$ lower [NV] and also narrower AO-PL contrast feature linewidths than sample S3-14N. $B_\perp$ values for maximum AO-PL contrast in (a, c) are shifted from 0 due to the presence of a background field.
}
\end{figure*}
\noindent

To compare AO-PL magnetic sensitivity between near-zero-field and low-field, we determine the linear slope extracted from the zero-crossing point of the central dispersive feature acquired with AO-PL lock-in detection for sample S3-14N at a fixed laser excitation power. Fig.~\ref{sup-fig:zeroLowField_slope} shows measurements and fits indicating that the slope $k$ at (b) low-field is $\sim3\times$ larger than that at (a) near-zero-field, despite lower overall AO-PL contrast at low-field. 

\begin{figure*}[!htb]
    \centering
    \includegraphics[width=0.75\textwidth]{Fig.18.pdf}
    \caption{\label{sup-fig:zeroLowField_slope}
    Demodulated AO-PL LIA spectra for sample S3-14N around (a) near-zero-field with fixed $B_\parallel = 0.05\,$mT and (b) $B_{\perp} \approx 1.06$\,mT with fixed $B_\parallel = $1.24\,mT. Linear fits are applied around the zero-crossing point. The slope $k$ from low-field measurements is larger than the near-zero-field slope by a factor of $\sim3\times$.
}
\end{figure*}
\noindent

\section{AO Sensitivity Characterization} \label{supp:sensitivity}
In this section, we detail the characterization of DC magnetic field sensitivity from our experimental setup using AO-PL LIA measurements. The PL signal collected by the photodiode is demodulated using a LIA with a sensitivity of 1\,mV, a modulation frequency of 100\,Hz, and a modulation depth $\approx 10\,\mu$T (see Appendix~\ref{supp:mod-calib}). The LIA time constant ($\tau$) is set to 300\,ms, with a filter order of 24\,dB per octave. 
\begin{figure}[!htb]
    \centering
    \includegraphics[width=0.48\textwidth]{Fig.19.pdf}
    \caption{\label{sup-fig:lockin_sensitivity}
    Demodulated AO-PL LIA spectrum for sample S2-15N around $B_{\perp} \approx 1.05$\,mT with fixed $B_\parallel = $1.24\,mT. A linear fit is applied around the zero-crossing point to convert the LIA output voltage to the magnetic field strength along the [$\bar{1}$10] direction. The sensitivity for magnetically sensitive (insensitive) regions of the spectrum is calculated using the standard deviation $\sigma$ from nine LIA output voltage measurements at $B_\perp \approx 1.05$\,mT ($B_\perp \approx 1.35$\,mT), respectively.
}
\end{figure}
Fig.~\ref{sup-fig:lockin_sensitivity} presents the demodulated AO-PL LIA spectrum around $B_\perp \approx 1.05$\,mT with fixed $B_\parallel = $1.24\,mT using sample S2-15N, averaged over nine measurements. The linear slope $k = 1228.685 \pm 18.973\,$V/mT, extracted from the zero-crossing point of the central dispersive feature, relates the LIA output voltage to the magnetic field strength along the [$\bar{1}$10] direction. The DC magnetic field sensitivity is calculated using the formula \(\frac{1}{k} \sigma  \sqrt{\tau}\), where $\sigma$ is the standard deviation of the nine AO-PL LIA voltage measurements at a specific magnetic field region of the spectrum, with results for magnetically sensitive (insensitive) regions (see Fig.~\ref{sup-fig:lockin_sensitivity}): $37.8 \pm 0.6$ ($5.3 \pm 0.1$)${\,\rm nT/\sqrt{Hz}}$, at $B_\perp \approx 1.05$\,mT ($B_\perp \approx 1.35$\,mT), respectively. 
For the central dispersive feature, the measured AO-PL contrast $C \approx 1.5\%$ and magnetic field linewidth $\Delta \gamma \approx 58\,{\rm \mu T}$ (equivalent to 1.37\,MHz, see Appendix~\ref{supp:linewidth_laser}) are comparable to those obtained with CW-ODMR for the NV-diamond samples and experimental conditions used in this study (see Appendices~\ref{supp:exp}, \ref{supp:linewidth_laser}).

% \bibliography{Weak-field}% Produces the bibliography via BibTeX.

\end{document}